\documentclass[12pt]{article}
\pdfoutput=1
\usepackage{amsmath,amsfonts,amssymb,amsthm,amssymb,bbm,bm}
\usepackage{pdfsync}
\usepackage{graphicx,import}
\usepackage[latin1]{inputenc}
\usepackage{hyperref}
\usepackage{empheq}
\usepackage[all]{xy}
\usepackage{stmaryrd}
\usepackage{rotating}
\usepackage{color}  
\usepackage{slashed}
\usepackage{cancel}
\usepackage{array, makecell} %
\usepackage{comment}
\usepackage{tikz-cd}
\usepackage{hhline}
\usepackage{comment}
\usepackage{cancel}
\usepackage{cite}
\makeatletter

\DeclareFontFamily{U}{matha}{\hyphenchar\font45}
\DeclareFontShape{U}{matha}{m}{n}{
    <5> <6> <7> <8> <9> <10> gen * matha
    <10.95> matha10 <12> <14.4> <17.28> <20.74> <24.88> matha12
     }{}
\DeclareSymbolFont{matha}{U}{matha}{m}{n}
\DeclareMathSymbol{\oright}       {2}{matha}{"69}

\newcommand{\doublehat}[1]{%
\begingroup%
  \let\macc@kerna\z@%
  \let\macc@kernb\z@%
  \let\macc@nucleus\@empty%
  \hat{\raisebox{.55ex}{\vphantom{\ensuremath{#1}}}\smash{\hat{#1}}}%
\endgroup%
}

\newcommand{\p}{\partial}

\newcommand{\bit}{\begin{itemize}}
\newcommand{\eit}{\end{itemize}}
\newcommand{\bd}{\begin{description}}
\newcommand{\ed}{\end{description}}

\newcommand{\bc}{\begin{center}}
\newcommand{\ec}{\end{center}}

\newcommand{\C}{{\mathbb C}}

\newcommand{\cM}{{\cal M}}
\newcommand{\cP}{{\cal P}}
\newcommand{\cJ}{{\cal J}}
\newcommand{\cN}{{\cal N}}
\newcommand{\cT}{{\cal T}}

\newcommand{\cL}{{\mathcal L}}

\newcommand{\cD}{{\mathcal D}}

\newcommand{\cS}{{\mathcal S}}
\newcommand{\cE}{{\mathcal E}}

\newcommand{\bN}{\bar{N}}
\newcommand{\cQ}{{\mathcal{Q}}}

\def\be#1\ee{\begin{align}#1\end{align}}
\newcommand{\bea}{\begin{eqnarray}}
\newcommand{\eea}{\end{eqnarray}}
\newcommand{\bs}{\begin{subequations}}
\newcommand{\es}{\end{subequations}}

\newcommand{\la}{\label}

\newcommand{\f}{\frac}

\newcommand{\bmm}{\bm{m}}
\newcommand{\bz}{{\bar{z}}}

\def\p{\partial}
\def\bD{D}

\def\d{\delta}

\def\rd{\mathrm{d}}
\def\pa{\partial }

\def\k{{\kappa^2} }

\newcommand{\tcM}{{\tilde{\mathcal M}}}

\newcommand{\scri}{\cal I}

\begin{document}

\begin{titlepage}

\unitlength = 1mm
\ \\
\vskip 3cm
\begin{center}

{\bf \Large{Sub-subleading Soft Graviton Theorem \\
from Asymptotic Einstein's Equations}\par}

\vspace{0.8cm}
Laurent Freidel$^1$, Daniele Pranzetti$^{1,2}$ and
Ana-Maria Raclariu$^1$
\vspace{1cm}

{\it $^1$Perimeter Institute for Theoretical Physics, 31 Caroline Street North, Waterloo, Ontario, Canada N2L 2Y5\\ \smallskip
\it $^2$ Universit\`a degli Studi di Udine, via Palladio 8,  I-33100 Udine, Italy}

\vspace{0.5cm}

\begin{abstract}

We identify in Einstein gravity an asymptotic spin-$2$ charge aspect whose conservation equation gives rise,  after quantization, to the  sub-subleading soft theorem.
Our treatment reveals that this spin-$2$ charge generates a non-local spacetime symmetry represented at null infinity  by pseudo-vector fields. Moreover, we demonstrate that the non-linear nature of Einstein's equations is reflected in the Ward identity through collinear corrections to the sub-subleading soft theorem. Our analysis also provides a unified treatment of  the universal soft theorems as conservation equations for the spin-0,-1,-2 canonical generators, while highlighting the important role played by the dual mass. 

\end{abstract}

\end{center}

\end{titlepage}

\tableofcontents

\section{Introduction}

It has been long known that the asymptotic symmetry group of gravity in four-dimensional asymptotically flat spacetimes (AFS) is infinite dimensional \cite{Bondi:1960jsa, BMS, Sachs62}. This enhancement remained largely overlooked until recently, when an unforeseen connection between the symmetries of null infinity and properties of scattering amplitudes in the infrared was uncovered. A prime example is the equivalence between Weinberg's soft graviton theorem \cite{Weinberg:1965nx}, the Ward identities associated with BMS supertranslation symmetry \cite{Strominger:2013jfa, He:2014laa} and the gravitational memory effect  \cite{ThorneB, Christodoulou:1991cr, Blanchet:1992br, Thorne:1992sdb, Favata:2010zu}. The latter was on the one hand identified with the spacetime Fourier version of the Weinberg soft pole, on the other hand related to transitions between the infinity of BMS vacua induced by gravitational flux \cite{Strominger:2014pwa}, rendering supertranslations physical.

Building on these ideas, the proposed extension of BMS to allow for local Lorentz transformations, or superrotations \cite{Barnich:2009se,  Barnich:2010eb, Barnich:2011mi, Barnich:2013axa}, led to the discovery of a new, subleading soft graviton theorem \cite{Cachazo:2014fwa, White:2014qia}. A derivation from the Ward identity associated with a Virasoro subgroup of the extended BMS group \cite{Kapec:2014opa}, as well as the identification of the corresponding observable, a new gravitational memory \cite{Pasterski:2015tva}, shortly followed. 
An equivalence between the subleading soft theorem and conservation laws was established by a further extension of the BMS group to include arbitrary smooth diffeomorphisms of the conformal sphere \cite{Campiglia:2014yka, Campiglia:2015yka}.
Moreover, a certain mode of the subleading soft graviton was shown to behave like the stress tensor of a two-dimensional conformal field theory \cite{Kapec:2016jld, Cheung:2016iub}, providing evidence for a dual description of gravity in 4D AFS in terms of a theory living on the celestial sphere \cite{deBoer:2003vf, Pasterski:2016qvg, Pasterski:2017kqt}.

A key lesson drawn is that the symmetries of gravity in four-dimensional asymptotically flat spacetimes are much richer than anticipated. The overarching goal of identifying the underlying symmetry structures and their implications has been approached with different methods. On the one hand, a reconsideration of the set of allowed boundary conditions and covariant phase space methods at null infinity \cite{Ashtekar:1981bq, Lee:1990nz} led to further extensions of the asymptotic symmetry group and their canonical analysis  \cite{Flanagan:2015pxa, Compere:2018ylh, Campiglia:2020qvc, Freidel:2021yqe}. On the other hand, the reformulation of the gravitational scattering problem in a basis of asymptotic boost eigenstates \cite{Pasterski:2016qvg, Pasterski:2017kqt} revealed the existence of an infinite tower of soft theorems \cite{Guevara:2019ypd, Guevara:2021abz}, which was remarkably shown to be governed by a higher-spin symmetry \cite{Strominger:2021lvk}. This symmetry was found to be perturbatively exact in self-dual gravity \cite{Ball:2021tmb} and, in the same context, explained via twistor methods in particular Penrose's non-linear graviton construction \cite{Adamo:2021lrv, Adamo:2021zpw}. At the classical level, the infinite tower of symmetries is a generic feature of any theory of gravity in 4D AFS, so it is natural to try to trace back its origin in the physically relevant case of Einstein gravity. 

In this paper, we take a first step in this direction by establishing an equivalence between the sub-subleading soft graviton theorem \cite{Cachazo:2014fwa, Zlotnikov:2014sva, Kalousios:2014uva, Luna:2016idw, Banerjee:2021cly} and the conservation law associated with a new class of asymptotic symmetries. Evidence for such a connection was previously provided in \cite{Campiglia:2016jdj, Campiglia:2016efb}, but the divergent behavior of the therein proposed vector fields at infinity and the lack of a well-defined symmetry action on the gravitational phase space precluded an identification of asymptotic symmetries.  {An interesting alternative approach was pursued  for linearized gravity in \cite{Conde:2016rom}, where the sub-subleading {soft graviton} theorem was {related} to {the} $1/r^2$ {contribution in a large-$r$ expansion} of the supertranslation charge.}

At the leading and subleading orders, the equivalence relies on a matching condition obeyed by the Bondi mass and angular momentum aspects, as well as their time evolution governed by constraint equations at leading order in a large-$r$ expansion \cite{He:2014laa, Kapec:2014opa, Campiglia:2014yka}. Our strategy is similar in spirit and builds on recent work \cite{Freidel:2021qpz} where symmetry considerations revealed the existence of a spin-2 charge with an evolution dictated by the remaining leading order evolution equation. This equation also follows from the Newman--Penrose  \cite{NP62, Newman:1962cia, Adamo:2009vu} analysis of asymptotic Einstein's equations \cite{Barnich:2011ty, Barnich:2013axa, Barnich:2016lyg, Barnich:2019vzx} where the spin-2 charge can be identified as the asymptotic value of the  Weyl scalar encoding incoming radiation. We demonstrate explicitly that, as in the leading and subleading cases, the associated asymptotic equation includes linear and quadratic components in the gravitational field. The former is identified with a sub-subleading soft graviton, while the canonical action of the latter on the gravitational phase space at null infinity is shown to be related to the sub-subleading soft graviton factor upon Fourier transform. Unlike the leading and subleading cases, there is also a cubic contribution which translates into a collinear correction to the sub-subleading soft theorem.

This paper is organized as follows. In Section \ref{sec:EOM} we review the results of \cite{Freidel:2021qpz} that revealed a clear  pattern organizing the asymptotic gravitational dynamics. In particular, we introduce the Weyl-BMS group (BMSW) \cite{Freidel:2021yqe} corresponding to the asymptotic limit to null infinity of the extended corner symmetry group of residual diffeomorphisms associated to a generic codimension-2 surface embedded in spacetime \cite{DonnellyFreidel,  Freidel:2020xyx, Freidel:2021cjp, Ciambelli:2021vnn}. This comprises all previously proposed BMS extensions \cite{Barnich:2009se,  Barnich:2010eb, Barnich:2011mi, Barnich:2013axa, Campiglia:2014yka, Compere:2018ylh}. Upon identifying the relevant physical quantities in terms of primary fields for the homogeneous subgroup of BMSW, the leading order asymptotic Einstein's equations are expressed compactly in terms of spin-weighted scalars. In Section \ref{sec:asy-cond} we specify fall-off conditions on the covariant phase-space variables which allow for the asymptotic evolution equations to be integrated. Finite corner charge aspects are constructed and associated with asymptotic symmetry generators.

The action of the corner charge aspects on the shear of null infinity is computed in Section \ref{sec:commutators}. In particular, the sub-subleading soft symmetry is shown to be generated by pseudo-vector fields quadratic in retarded time.  In Section \ref{cl-st} we use these actions to demonstrate how the conservation laws for the renormalized corner charge aspects imply the three soft theorems. At the sub-subleading order we show that the structure of the asymptotic Einstein's equation associated with the spin-2 charge yields higher order corrections in $G_N$ to the factorization properties of tree-level scattering amplitudes. These corrections take the form of new collinear terms in the sub-subleading soft factor which are discussed in Section \ref{sec:corrections}. Concluding remarks are presented in Section \ref{sec:conc} and technical details are collected in Appendices \ref{confD}, \ref{app:comm}, \ref{app:conventions} and \ref{osc}.

\section{Asymptotic equations of motion}\la{sec:EOM}

Introducing  the retarded Bondi coordinates $x^\mu=(u,r,\sigma^A)$,
where $\sigma^A$ denote coordinates on the celestial 2-sphere,
asymptotically flat metrics in the Bondi gauge near future null infinity ($\mathcal{I}^+$) take the general form \cite{Barnich:2009se, Barnich:2010eb}
\be \label{ds}
\rd s^2 = -2 e^{2\beta} \rd u (\rd r + \Phi \rd u) + r^2 \gamma_{AB} \left(\rd\sigma^A - \frac{\Psi^A}{r^2}\rd u \right)\left(\rd\sigma^B - \frac{\Psi^B}{r^2} \rd u\right).
\ee
In a large-$r$ expansion, generic solutions to the asymptotic Einstein's equations to order $r^{-1}$ are of the form
\begin{subequations}\la{eq:FallOff}
\begin{align}
\Phi&= \frac{R({q})}4 - \frac{  M}{r}+o\left(r^{-1}\right)\,,\\
\beta&=-\frac{1}{32 } \frac{  C_{AB} C^{AB}}{r^2}+o\left(r^{-2}\right)\,,\\
\Psi^A&=-\frac12 {D}_B C^{BA}- 
\frac{1}{  r } 
\left(\f23 \cP^A- \frac12 C^{AB} {D}^C C_{CB} - \frac{1}{16 } \pa^A  \left(C_{BC} C^{BC}\right) \right) +o\left(r^{-1}\right)\,,\label{PA}\\
\gamma_{AB}&=  {q}_{AB}  + \frac{1}{r}C_{AB} + \frac{1}{ 4r^2} q_{AB}\left(C_{CD}C^{CD} \right)
+ \f1{r^3}\left({ {\f13}}{\cT}_{AB} +\f1{16}  C_{AB} (C_{CD} C^{CD})\right)  +  o\left(r^{-3}\right)\,,
\la{gamma}
\end{align}
\end{subequations}
where $\pa_u q_{AB}=0$. All indices are raised with the inverse sphere metric\footnote{We do not assume that $q_{AB} $ is the round sphere metric. } $q^{AB}$ and the shear  $C_{AB}$ is symmetric and traceless. 

The asymptotic symmetry group is the Weyl-BMS group (BMSW) \cite{Freidel:2021yqe} generated by vector fields that act on $\scri^{+}$ as 
\be 
\xi_{(T,Y,W)}= T\pa_u + Y^A\pa_A + W (u\pa_u- r\pa_r).
\ee
These are parameterized by a vector field $Y^A(\sigma^A)$ and functions $W(\sigma^A), T(\sigma^A)$ on the sphere.
BMSW contains a  homogeneous subgroup  ${H}_S:=(\mathrm{Diff}(S)\ltimes \mathrm{Weyl})$ acting on  the normal subgroup of supertranslations. 
The subgroup of BMSW implemented canonically on the asymptotic phase space  is the generalized BMS group (GBMS) \cite{Campiglia:2015yka, Compere:2018ylh,Campiglia:2020qvc} which preserves the measure $\sqrt{q}$. The metric and its determinant transform under BMSW as 
\be\label{delq}
\delta_{(T,Y,W)} q_{AB}= (\cL_Y -2 W) q_{AB}, \qquad \delta_{(T,Y,W)} \sqrt{q} = D_A Y^A - 2 W,
\ee 
hence upon restricting to GBMS the Weyl factor becomes $W_Y=\frac12 D_A Y^A$. 

The homogeneous subgroup $H_S$ can be used to organize the metric components in terms of primary fields for $H_S$ \cite{Freidel:2021qpz}.  These are denoted by $O_{(\Delta,s)}$ and are labelled by a spin $s$ and a conformal  dimension $\Delta$ \cite{ Barnich:2011ty, Barnich:2019vzx}. By definition, for a particular cut $u=0$ of $\scri^+$, a primary field transforms \emph{homogeneously}
under   $H_S$ as
\be\la{eq:prim-trans} 
\delta_{(Y,W)} O_{(\Delta,s)}
= (\cL_Y + (\Delta-s) W ) O_{(\Delta,s)},
\ee
with $\cL_Y$ the Lie derivative along $Y$. 
We assign spin $+1$ to vector fields $\pa_A$, spin $+2$ to $\pa_{\langle A}\pa_{B\rangle}$, etc. and spin $-1$ to dual forms $\rd \sigma^A$, as well as dimension $\Delta = 1$ to $\p_u$ and $r$. In this convention,  positive spin operators correspond to symmetric traceless forms while negative spin ones correspond to symmetric traceless tensors, which we denote by
\be 
O_{(\Delta,s)}:= O_{\Delta \langle A_1 \cdots A_s\rangle},\qquad  O_{(\Delta,-s)}:= O_{\Delta}^{ \langle A_1 \cdots A_s\rangle}, \quad s \geq 0.
\ee 
{This parallels the re-organization of asymptotic data in a conformal primary basis \cite{Pasterski:2016qvg, Pasterski:2017kqt, Pasterski:2021fjn, Pasterski:2021dqe}}. On the space of homogeneous primary fields we can perform certain operations that map primaries onto primaries. A basic example of such an operation is the  metric contraction 
\bea
q_{AB}: (\Delta,s) &\to& (\Delta, s+2),
\eea
which raises\footnote{This is consistent with our definition \eqref{eq:prim-trans} and the transformation \eqref{delq}.} the spin by $2$.
Similarly, contraction with the inverse metric lowers the spin by $2$.   Another candidate operation is the contraction with the Levi-Civita derivative $D_A$ which raises the spin by 1.\footnote{\label{Daction}Its action is explicitly given by 
\be 
D_A:O^{A_1\cdots A_s} \to D_{A_s}O^{A_1\cdots A_s}, 
\qquad
D_A:O_{A_1\cdots A_s} \to D_{\langle A_{s+1}}O_{A_1\cdots A_s \rangle },
\ee} This operation does not preserve the primary condition; however, upon introducing a conformal connection $\Upsilon_A$, one can construct a conformally invariant derivative (see Appendix \ref{confD})
\be 
\cD_A O_{(\Delta,s)} = (D_A +(\Delta + s) \Upsilon_A)O_{(\Delta,s)},
\ee 
where the action of $\cD_A$ is contracted or symmetrized as in footnote \ref{Daction}.
This maps primaries onto primaries with shifted dimension and spin,
\be 
\cD_A: (\Delta,s) \to (\Delta+1,s+1).
\ee 
The connection $\Upsilon_A$ transforms as a one form under diffeomorphisms and inhomogeneously under rescalings, namely
\be \la{dUp}
\delta_{(Y,W)} \Upsilon_A = \cL_Y\Upsilon_A  - D_A W. 
\ee 
There is a unique connection $\Upsilon_A(q)$ with the property that $\Upsilon_A(\mathring{q})=0$ for the round sphere  metric $\mathring{q}.$ 
The importance of the conformal derivative to the canonical analysis of GBMS was first revealed by Campiglia and Peraza in \cite{Campiglia:2020qvc}. It was also used by Donnay and Ruzziconi in \cite{Donnay:2021wrk}\footnote{We thank S. Pasterski for bringing this to our attention.} and it is a central feature of the holographic fluid perspective developed by Ciambelli et al. \cite{Ciambelli:2019lap, Ciambelli:2019bzz}. 

Another candidate operation on primaries is the time derivative $\pa_u: (\Delta,s) \to (\Delta+1, s)$ which raises the dimension and preserves the spin. It turns out that if $O_{(\Delta,s)}$ is a homogeneous primary, then $\pa_u O_{(\Delta,s)}$ is a primary if and only if it transforms covariantly under supertranslations \cite{Donnay:2021wrk, Freidel:2021qpz}. For instance, the shear $C_{AB}$ is a primary of dimension-spin $(1,2)$, while its time derivative $N^{AB}:=\dot{C}^{AB}$ is not. However, the news tensor $\hat{N}^{AB}$ defined by \cite{Geroch:1977jn, Compere:2018ylh}
\be
\label{vac-N}
\hat{N}^{AB}:= N^{AB}- \tau^{AB},\qquad 
\tau_{AB}:=2(D_{\langle A}\Upsilon_{B \rangle} + \Upsilon_{\langle A} \Upsilon_{B\rangle} ),
\ee 
is a primary of dimension-spin $(2,-2)$ (see Appendix \ref{confD}). $\tau_{AB}$ is the so-called Liouville or Geroch tensor \cite{Geroch:1977jn, Compere:2018ylh}.
Moreover, the time derivative of the news $\cN^{AB} := \p_u \hat{N}^{AB}$ is a primary of dimension-spin $(3,-2)$. More generally $\pa_u^n\cN^{AB}$ is a primary of $(\Delta, s) = (n+3,-2)$. 

 Building on these ideas, primary fields consisting of components of the metric in a large-$r$ expansion to order $r^{-1}$ were identified in \cite{Freidel:2021qpz}. 
The list of primaries additionally includes the energy current $\cJ^A$ (3,-1), the covariant mass $\cM$ (3,0), the covariant dual mass $\tilde{\cM}$ (3,0), the momentum $\mathcal{P}_A~ (3,1)$ and the spin-$2$ tensor $\cT_{AB}$ (3,2).
The momentum and spin-$2$ field already appear in the metric expansion \eqref{eq:FallOff}, while the energy current and covariant masses take the form
\be
\cJ^A &:= \frac12 D_B N^{AB} + \frac14 \pa^A R(q)\,, \la{JA}\\
{\cM}&:= M + \f18 N^{AB} C_{AB}\,, \la{real-mass-aspect}\\
\tcM&:= \f14   \epsilon^A{}_C \left({D}_{A} \bD_B{ C^{CB}} +  \f12  {N}^{CB}C_{AB}\right)\, \la{dual-mass-aspect}.
\ee
 Here $\epsilon_A{}^B$ is the complex structure on the 2-sphere defined through the volume form $\epsilon_{AB}$, namely
\be
\epsilon_A{}^B := \epsilon_{AC} q^{CB},
\qquad 
\epsilon_A{}^B\epsilon_B{}^C =-\delta_A^C\,.
\ee
The relevant primary fields and their weights are summarized in Table \ref{tab:semi-primaries}.
\begin{table}[ht] 
\centering
\begin{tabular}{|c|c| c| c | c| c| c|c|c|c|} 
 \hline
 Primary Fields & $C_{AB}$ & $\cN^{AB}$ & $\cJ^A$ & $\cM$ & $\tilde{\cM}$ & ${\cP}_A$ & ${\cT}_{AB}$ \\ 
 \hline 
Dimension-Spin $(\Delta, s)$   &  (1,2)  &  (3,-2)  & (3,-1) & (3,0) & (3,0) &(3,1) &(3,2)  \\
 \hline
 \end{tabular}
 \caption{Fields transforming covariantly according to \eqref{eq:prim-trans} under the homogeneous component of BMSW.}
 \la{tab:semi-primaries}
 \end{table}

Under supertranslations, the quantities in Table \ref{tab:semi-primaries} acquire inhomogeneous shifts. For example,
\be
 \delta_{T} {\cal J}^A &=
T\pa_u {\cJ}^A+ \frac{1}{2}\cN^{AB} \pa_B T,\\
\delta_{T} {\cal M} &= T\pa_u {\cal M}+ \cJ^A \pa_A T\,,\\
\delta_T {\cal P}_A&=  T\pa_u  {\cal P}_A 
 + 3 \left({\cal M}\pa_AT  +   \tilde{\cM} \tilde\pa_AT\right).
\ee
Remarkably, Einstein's equations can be reconstructed by identifying the combinations of fields and derivatives that transform homogeneously under arbitrary BMSW transformations \cite{Freidel:2021qpz}. We illustrate this in a short example. It can be easily shown that there are no translationally covariant scalar combinations at dimension $3$. At $(\Delta, s) = (4,0)$ the set of all parity-even primaries constructed from primary fields is\footnote{Note that products of primaries are also primary.} 
\be 
C_{AB} \cN^{AB},\quad \dot{\cM}-\f12 D_A \cJ^A.  
\ee
It turns out that the unique linear combination transforming homogeneously under supertranslations is
\be 
\cE:= \dot{\cM} - \frac12 D_A \cJ^A -\frac18 C_{AB} \cN^{AB}.
\ee 
In the absence of sources, the only covariant equation is therefore $\cE=0$. This is one of the asymptotic vacuum Einstein's equations.
We can continue this exercise for different spins and parity conditions. All supertranslation primaries that can be constructed from the asymptotic metric expansion \eqref{eq:FallOff} have dimension $\Delta = 4$ and spin $s=-2,-1,0,1,2$, yielding all equations of motion to the same order in a large-$r$ expansion.

The equations of motion can be compactly written by introducing a holomorphic frame $m= m^A\pa_A$ with coframe $\bmm = m_A \rd \sigma^A$
and normalization $m^A\bar{m}_A=1$. In terms of these frame fields, the sphere metric $q_{AB}$ and the volume form $\epsilon_{AB}$ are given by\footnote{In complex coordinates, the normalization implies that for the round sphere $\bmm = P \rd \bz$,
where $P:= \frac{\sqrt{2}}{(1+z\bar{z})}$. We do not restrict to the round sphere metric case and do not fix the form of $\bmm$.} 
\be
\label{trm}
q_{AB} =  (m_A \bar{m}_B + m_B \bar{m}_A)\,,\qquad
\epsilon_{AB} = {-}i( m_A \bar{m}_B - m_B \bar{m}_A)\,. 
\ee
Both $m_A$ and $\bar{m}_A$ have $(\Delta, s) = (0,1)$. They are however distinguished by their \emph{helicity} (also called spin-weight when the  metric is spherical \cite{Pasterski:2020pdk, Pasterski:2021fjn, Pasterski:2021dqe}): $m_A$ has helicity $+1$ while $\bar{m}_A$ has helicity $-1$. 
We can use the frame field to convert spin-$s$ tensors into scalars of a given helicity. By convention, positive and negative helicity scalars can be obtained by contraction with  $m^A$ and $\bar{m}_A$ respectively, 
\be 
O_s = O_{A_1\cdots A_s} m^{A_1}\cdots m^{A_s},
\qquad 
O_{-s} = O^{A_1\cdots A_s} \bar{m}_{A_1}\cdots  \bar{m}_{A_s}.
\ee 
This implies that $O_{-s}= \bar{O}_{s}$, meaning that negative helicity scalars are complex conjugates of positive helicity ones.

Given  the phase space variables  $( \cN^{AB},\cJ^A, \cM,\tcM,\cP_A,\cT_{AB})$ of conformal dimension $3$, we define 
the following  (spin-weighted) scalars
\be
&C:= C_{AB} m^A m^B, \qquad \cN:=\cN^{AB}\bar{m}_A\bar{m}_B,\qquad \cJ:= \cJ^A\bar{m}_A,\cr
&\cM_{\C} := \cM+i\tcM,\qquad
\cP:= \cP_A {m}^A,
\qquad
\cT:= \cT_{AB} {m}^A{m}^B\,.
\ee
We have introduced the complex mass $\mathcal{M}_{\mathbb{C}}$ which is a complex linear combination of the mass and its dual.
In spherical complex coordinates this definition implies that $ C=P^{-2} C_{zz}$ and
$N = P^2 N^{zz}= P^{-2} N_{\bar{z}\bar{z}}$ in agreement with the standard convention that (outgoing) positive and negative helicities correspond to holomorphic and anti-holomorphic forms respectively \cite{He:2014laa, Kapec:2014opa}.

We denote $D = m^A D_A$ the Cartan derivative\footnote{\la{f1}This is such that 
\be 
\la{D-def}
DO_s = m^A m^{A_1}\cdots m^{A_s} D_A O_{A_1 \cdots A_s} =  (D_m -is \Omega)O_s,
\ee where $D_m:= m^A D_A$, with $D_A$ the covariant derivative and where $i\Omega := \bar{m}_B D_m m^B = \bar{m}_B D_{\bar{m}} m^B $ is the 2d spin connection.} along $m^A$ and $\bar{D} = \bar{m}_A D^A$ the Cartan derivative along $\bar{m}$ (see Appendix \ref{SpinC}).
$D$ raises the spin weight by $1$, while $\bar{D}= \bar{m}_A D^A $ lowers it by $1$.
Upon contraction with the frame field, \eqref{JA} can then be recast as
\be
\cJ= \frac12 D N +\frac14 \bar{D} R\,,\la{J}
\ee
or equivalently in terms of the contraction $\hat{N} = \hat{N}^{AB}\bar{m}_A \bar{m}_B $ of the news tensor \eqref{vac-N},
\be
\cJ = \frac{1}{2}D\hat{N},
\ee
where we have used (see Appendix \ref{comm-curv}) $\frac{1}{2}\bar{D}R = -D\bar{\tau}$, with $\bar{\tau} := \bar{m}_A \bar{m}_B \tau^{AB}$ and $N:= \bar{m}_A\bar{m}_B N^{AB}$. 
The asymptotic evolution equations \cite{Barnich:2011mi,Nichols:2018qac, Freidel:2021qpz} can then be compactly expressed as
\begin{subequations}\label{eom}
\be
\dot{\cJ} &=\tfrac12  D \cN\,,\la{cJ}\\
\dot{\cM}_{ \C}&= D \cJ + \tfrac14 C \cN\,, \la{cM}\\
\dot{\cP}&= D\cM_{ \C} + C \cJ\,, \la{cP}\\
\dot{ \cT}&= D \cP + \tfrac32 C\cM_{ \C}\,,\la{cT}
\ee
\end{subequations}
and their complex conjugates.

It will prove convenient to organize  aspects of  conformal dimension $3$ in terms of their helicity and denote 
 \be\cQ_{-2}:=\frac{\cN}{2},\qquad  \cQ_{-1}:= \cJ,\qquad \cQ_0:=\cM_\C,\qquad \cQ_1:=\cP,\qquad \cQ_2:=\cT,
 \ee
allowing for the asymptotic Einstein's equations to be compactly expressed as
\be
\dot{\cQ}_{s} = D \cQ_{s-1} + \frac{(1+s)}{2} C \cQ_{s-2}\,,
\ee
for $s=-1,0,1,2 $. This analysis can be repeated near $\mathcal{I}^-.$

We conclude this preliminary section with a clarifying remark. In this paper, we emphasize the corner and celestial fluid perspective \cite{Freidel:2019ees, Freidel:2019ofr, Freidel:2020xyx,Donnelly:2020xgu, Penna:2015gza, Penna:2017bdn, Ciambelli:2019lap, Ciambelli:2019bzz} where $\cM_{\mathbb{C}}$, $\cP$ and $\cJ$ are the complex energy density, momentum and energy current of the celestial fluid. From the gravity point of view, $\cM$ and  $\cP$ are the momentum and angular momentum aspects. The distinction between bulk and boundary stems from the fact that translations on the celestial sphere arise from bulk rotations. We find the holographic point of view on the asymptotic dynamics quite powerful and inspiring. Of course, this is just a change of perspective and nomenclature with respect to the standard relativist's bulk point of view where $\cM$ is sometimes denoted $\cP$, while $\cP$ is denoted $\cJ$, such as in \cite{Barnich:2021dta}. We hope that this doesn't introduce confusion.

\subsection{Asymptotic conditions and integrated charges}
\label{sec:asy-cond}

The Einstein constraint equations played an important role in establishing the equivalence between asymptotic symmetries and soft theorems at the leading \cite{Strominger:2013jfa,He:2014laa} and subleading orders \cite{Kapec:2014opa, Campiglia:2014yka, Campiglia:2015yka}. In order to integrate the asymptotic evolution equations \eqref{eom} and generalize the analysis to the sub-subleading case, the asymptotic behavior of the dressed news and the charges at large retarded times needs to be specified.
As we will see, to access the sub-subleading soft theorem one must impose that $\hat{N} = O(|u|^{-\alpha})$ where $\alpha > 3$. For the leading and subleading soft theorems weaker fall-offs are sufficient, namely $\alpha >1$ \cite{Strominger:2013jfa}  and $\alpha>2 $ \cite{Campiglia:2020qvc} respectively. In order to avoid logarithmic corrections \cite{Laddha:2017ygw,Laddha:2018myi}, $\alpha \notin \mathbb{N}$ will be assumed throughout.

These fall-offs on the news are necessary to ensure that all generators of asymptotic symmetries $(\cJ,\cM_\C,\cP,\cT)$ decay to zero at $\scri^+_+$, where the geometry reverts to a radiative vacuum. Specifically, 
\be \label{limQ}
 \lim_{u\to + \infty} \cQ_s(u,z) =0\,,
 \ee
 for $s=-2,-1,0,1,2$. Here and henceforth, arguments $z$ compactly denote dependence on the transverse coordinates $z, \bz.$
Note that it is  essential to use the covariant charge aspects to be able to impose boundary conditions that capture the soft physics. While these asymptotic conditions are restrictive,\footnote{For instance, imposing \eqref{limQ} for $\cM_\C$ excludes the presence of black holes.} they are well adapted to the S-matrix context. In particular, \eqref{limQ} allows one to define the charge aspects as integrals over their flux
 \be 
 \label{leading-flux}
 \cM_\C (u,z) =\int_{+\infty}^u \rd u' \dot{\cM}_\C (u',z),
 \ee and similarly for $\cP$ and $\cT$.  Subject to the asymptotic fall-offs of $\hat{N}$, we deduce that
 \be\la{alpha}
 \cQ_s = O(u^{1+s-\alpha}) \quad {\rm when}\quad u\to +\infty\,.
 \ee
Consequently, the condition $\alpha>3$ is necessary to integrate the spin $s = 2$ charge $\cT$. Note that these conditions do not fix the value of $C(u,z)$ when $u\to +\infty$. Nevertheless, this value can be set to $0$ by performing a combination of supertranslation, dual supertranslation and superrotation which amounts to fixing the asymptotic frame of reference at timelike infinity to be a center of mass frame. We henceforth assume that $C= O(u^{-\alpha +1})$ 
at $u=+\infty$.
Of course, such a choice cannot be independently made at $u=-\infty$ as the asymptotic value of $C$ is determined by the memory effect.
 
These fall-off conditions are not sufficient to ensure that $\cP$ and $\cT$ have finite limits when $u\to -\infty$.
 To remedy this problem, following \cite{Freidel:2021qpz} one can consider a non-radiative phase space defined by the conditions\footnote{These conditions are equivalent to the typically employed ones for the non-radiative phase space, namely $\hat{N} = 0$. }
\be
\cN=0=\cJ\,.\la{NR}
\ee
In this case, the shear is simply given by its memory components
\be 
C_{\mathrm{M}}(u,z):= c(z) + u \tau(z).
\ee 
The non-radiative corner phase space is then parameterized by the renormalized corner charge aspects  \cite{Freidel:2021qpz}
\be
\hat{m}_\C= \f8{\kappa^2}\cM_\C\,,\quad \hat{p}= \f8{\kappa^2}\f12\left(\cP-u D \cM_\C\right)\,\quad \hat{t}= \f8{\kappa^2} \f13\left(\cT -uD\cP + \left(\f{u^2}2 D^2-\f32 \left(\int^u \!\! C\right)\right)\cM_\C\right)
 \,,\la{corq}
\ee
where $\kappa=\sqrt{32\pi G}$. These are \emph{time independent} when \eqref{NR} holds. 
The reason for the overall numerical rescaling   will become clear when we compute their action on $C$ in Section \ref{sec:commutators}.
The first two aspects $(\hat{m}_{\C},\hat{p})$ define a moment map for the generalized BMS group (this was proven for the case $\tcM=0$ in \cite{Barnich:2021dta}). Additionally, the spin-$2$ charge $\hat{t}$ defines a moment map for an extension of the generalized BMS group at null infinity to include the spin-2 charge aspect $\hat{t}$ \cite{FMP}.

To summarize, in order to establish the soft theorems, one needs to impose appropriate boundary conditions such that the renormalized charge aspects vanish at $\mathcal{I}^+_+$,
\be
\lim_{u\to +\infty} \hat{q}_s(u,z)=0\,
\ee
and are finite 
at $\mathcal{I}^+_-$, namely
\be 
\label{charge-aspects}
\lim_{u\to -\infty} (\hat{m}_{\C},\hat{p},\hat{t})(u,z)= ({m}_{\mathbb{C}}(z),{p}(z),{t}(z))\,.
\ee 
The asymptotic symmetry generators can then be compactly expressed as
\be
Q_{(T, Y, Z)}:=\int \rd^2 z \sqrt{q} \left(Tm_\C+ Yp+Zt\right)( z)\,.\la{SQ}
\ee

In order to compute the symmetry action on $C$ it will be necessary to consider the integrated\footnote{The boundary condition 
 $ \cQ_s = O(u^{1+s-\alpha})$ at $u\to +\infty$ allows one to do so.} asymptotic equations of motion
\be
\cM_{\C}(u) & = \frac12 D^2 (\pa_u^{-1}\hat{N})  + \tfrac14 \pa_u^{-1}(C \cN)\,, \label{M}\\
\cP(u)  
&=\frac12 D^3   (\pa_u^{-2}\hat N) + \tfrac14  D \pa_u^{-2}(C \cN)  + \pa_u^{-1}(C \cJ)\,,\label{P}\\
\cT(u)  &
= \frac12 D^4   (\pa_u^{-3} \hat N) + \tfrac14  D^2 (\pa_u^{-3}C \cN)  + D \pa_u^{-2}(C \cJ) + \frac32 \pa_u^{-1}(C \cM_\C)\,,\label{T}
\ee
where we introduced the symbolic notation 
\be
 (\pa_u^{-n} \hat N)(u) := \int_{+\infty}^{u}\rd u_1 \int_{+\infty}^{u_1} \rd u_2 \cdots \int_{+\infty}^{u_{n-1}} \rd u_n\, \hat N(u_n) .\label{iiint}
\ee
This notation will come in handy in the next sections. An identity that will be essential to our story is the Leibniz rule for the pseudo-differential operator $\pa_u^{-1}$  \cite{Gelfand-Dickey, Adler, DickeyB, Bakas:1989um, Bonora:1994fq}
\be\la{Leib}
\pa_u^{-1}(PF)= \sum_{n=0}^{\infty} (-1)^n (\pa_u^n P) (\pa_u^{-n-1}F).
\ee
The sum truncates when $P$ is a polynomial in $u$.\footnote{More details on pseudo-differential calculus are provided in Section \ref{sec:c-subsub}.} Note that the asymptotic conditions \eqref{alpha} ensure that the integrals \eqref{M}-\eqref{T} are well defined.
The order of integral labels in \eqref{iiint} is tailored to the choice of boundary conditions  at $\scri^+_+$, corresponding to the no-radiation condition \eqref{limQ} at $u=+\infty$.

 \subsection{Brackets}

The basic bracket needed to compute the action of the charges \eqref{SQ} on the asymptotic shear is \cite{He:2014laa, Ashtekar:1978zz, Ashtekar:1981sf, Ashtekar:2018lor} 
 \be
 \label{NCbracket}
 \{  \hat N(u,z), C(u', z')\}&= \f \k 2\delta(u-u') \delta(z,z')\,.
 \ee
 Note that  on $\scri^+$, it is the shifted news \eqref{vac-N} that is canonically conjugate to the shear $C$. 
The Poisson bracket \eqref{NCbracket} implies the following  other brackets\footnote{We are assuming that $\{m^A, C\} = 0$ which may have to be revisited for extensions of BMS beyond Virasoro \cite{Compere:2018ylh, Freidel:2021yqe}.}
\be
\{ \cN(u,z), C(u', z')\}&= \f{\kappa^2}2 \pa_u\delta(u-u') \delta(z,z')\,,\la{NC}\\
\{ \cJ(u,z),C(u', z')\} &= \f{\kappa^2}4 \delta(u-u') D_z\delta(z,z')\, ,\la{JC}\\
\{ C(u,z), C(u', z')\} &= 0\, .
\ee

Quantum commutators are simply obtained by defining $[\cdot, \cdot ] = -i\hbar\{\cdot,\cdot\}$.
At the quantum level we therefore have 
\be \label{NCC}
[ \cN(u,z), C(u', z') ] &= {-i} \f\k2  \pa_u\delta(u-u') \delta(z,z')\,.
\ee
 The delta function on the sphere is dual to the measure 
 $\epsilon = i P^2 \rd z\wedge \rd \bar{z}$, meaning that 
 \be
 \delta(z,z') = \frac{\delta^{(2)}(z-z')}{P^2}.
 \ee

\section{Charge action}
\label{sec:commutators}

In this section we evaluate the action of the local charges \eqref{corq} on the shear $C$. The commutation relations follow straightforwardly from
\eqref{NC} and \eqref{JC} and are explicitly computed in Appendix \ref{app:comm}. For clarity, we continue working on $\mathcal{I}^+$, but a similar analysis pertains to $\mathcal{I}^-$ and will be discussed in Section \ref{cl-st}.

\subsection{Leading action}
We start with the complex mass aspect \eqref{M} which can be split as \cite{Strominger:2013jfa, He:2014laa}
\be\la{DM}
\cM_\C= \cM_{\mathrm{S}} +\cM_{\mathrm{H}}\,,
\ee
where $\cM_{\mathrm{S}}$ is linear in $C$, while $\cM_{\mathrm{H}}$ is quadratic in $C,\bar{C}$. These are explicitly given by
\be
\cM_{\mathrm{S}}(u,z)& =  \int_{+\infty}^u \rd u'  D \cJ(u',z)\,,\\
\cM_{\mathrm{H}}(u,z) &=  \frac14 \int_{+\infty}^u \rd u'   C(u',z) \cN(u',z)\,.
\ee
A related split, and the transformation properties of its soft and hard components under asymptotic symmetries were studied in \cite{Donnay:2021wrk}.
The brackets \eqref{NC} and \eqref{JC} allow us to evaluate
\be
\{\cM_{\mathrm{S}}(u,z), C(u', z')\}
&= -\f{ \kappa^2}4\theta(u'-u) D_z^2\delta(z, z')\,,\label{MsC2}
\\
\{ \cM_{\mathrm{H}}(u,z), C(u', z')\}
&=\f{\kappa^2}8\pa_{u'}[ C(u',z) \theta(u'-u) ] \delta(z, z')\,,\label{MhC}
\ee
where $\theta(x)$ is the unit step function defined in \eqref{theta-fctn}. Here and in the following sections, the derivatives {are as defined in \eqref{D-def}} and in particular do not carry any spatial index. The subscripts $z, z'$ are simply introduced to keep track of the variables they act on.  

Putting these together, we find that for $u' >u$ 
\be
\f8{\kappa^2}\{ \cM_\C(z,u), C(u', z')\}
=  \left(\pa_{u'} C(u',z)  - 2 D_z^2\right) \delta(z, z'),
\label{CM}
\ee
or equivalently, 
in terms of the supertranslation corner aspect defined by \eqref{charge-aspects}
\be\la{mC}
\boxed{\,\,
\{ m_\C(z), C(u', z')\}
=  \left(\pa_{u'} C(u',z')   - 2D_z^2\right) \delta(z, z')\,.
\,\,
}
\ee
According to \eqref{SQ}, the supertranslation charge\footnote{From now on we use the shortcut notation $\int_S F := \int_S \rd^2 z \sqrt{q} F.$}
 \be
Q_T :=  \int_S  T(z) m_\C( z)\,
 \ee
 induces the following symmetry transformation on $C$
 \be
\delta^0_T C(u, z):=  \{Q_T,  C(u, z)\} = T\pa_{u} C(u,z) - 2D_z^2 T\,,
 \ee
 which is the usual action of supertranslations on the shear. The homogeneous component of the transformation indicates that the charge $Q_T$ is associated with a vector field
 \be
 \xi_T=T\p_u\,.
 \ee

It is worth emphasizing that it is the charge constructed from the complex mass aspect that reproduces the expected Lie derivative action on $C$. On the other hand, in order for the ``standard'' supertranslation charges associated with the real mass aspect \eqref{real-mass-aspect} to act correctly, one ought to impose an equivalence relation on phase space \cite{Strominger:2013jfa, He:2014laa}, 
\be 
\label{C-constraint}
\tilde{\mathcal{M}} \propto \left[\left(\bar{D}^2-\f12 N\right) C-\left(D^2 -\f12 \bar{N}\right)\bar{C} \right]_{\mathcal{I}^{+}_{\mp}} = 0.
\ee 
One has to be careful with such constraints because they are second class, meaning $\{\tilde{\mathcal{M}}, C \} \neq 0$.
To further clarify this, we consider the action of the anti-holomorphic charges on the holomorphic shear $C$,
\be 
\{\bar{Q}_T, C(u, z)\} = T \p_{u} C(u, z).
\ee 
The real and imaginary parts of the (complex) supertranslation charge then act respectively as
\be 
\begin{split}
\left\{\frac{Q_T + \bar{Q}_T}{2}, C(u,z) \right\} &= T \pa_{u} C(u,z)   - D_z^2 T,\\
\left\{\frac{Q_T - \bar{Q}_T}{2}, C(u,z) \right\} &=  - D_z^2 T.
\end{split}
\ee
We see explicitly that the dual mass contains a purely soft contribution, providing a complementary perspective  on the mysterious factor of 2 mismatch for the real charge action first observed and remedied in \cite{He:2014laa}. While the importance of the dual mass was previously pointed out in \cite{Godazgar:2018dvh,Godazgar:2019dkh, Godazgar:2020kqd}, its interpretation in a scattering context remains elusive. We leave a complete understanding of these interesting issues to future work.

 \subsection{Subleading action} 

The momentum aspect is given by
\be
\cP(u,z)=   \int_{+\infty}^u \rd u'  \dot{\cP}(u',z)\,.
\ee
Using \eqref{P} this can be decomposed as
\be\la{DP}
\cP= \cP_{\mathrm{SS}} +\cP_{\mathrm{SH}}+\cP_{\mathrm{HH}}\,,
\ee where the subscripts label the three terms in \eqref{P}, namely $\cP_{\rm HH}$ is the hard component associated with $ C\cJ$,  $\cP_{\rm SH}$ corresponds to $D\cM_{\mathrm{H}}$, while  $\cP_{\rm SS}$ is the soft contribution $D\cM_{\rm S}$. 
The brackets of the individual terms with $C$ are computed in Appendix \ref{app:subleading-charges} and the results take the form
\be
\{ \cP_{\mathrm{SS}}(u,z), C(u', z')\} 
&={{\frac{\kappa^2}{4}}}(u'-u) \theta(u'-u)  D_z^3\delta(z, z')\,,
\\
\{ \cP_{\mathrm{SH}}(u,z), C(u', z')\}
&={{-\frac{\kappa^2}{8}}}D_z\pa_{u'}[ C(u',z) \theta(u'-u)\delta(z, z')(u' - u) ]\,,
\\
\{ \cP_{\mathrm{HH}}(u,z), C(u', z')\}
&=-{{\frac{\kappa^2}{4}}}[ C(u',z) \theta(u'-u) ] D_z\delta(z, z')\,.
\ee
Overall, these imply that for $u' >u$ 
\be 
\{ \cP(u,z), C(u', z')\} &=-\f{\kappa^2}8 (u'-u) D_z[\left(\pa_{u'}C(u',z) - 2 D_z^2\right)\delta(z, z')]\cr
&- \f{\kappa^2}8 \left(D_{z}[C(u',z)\delta(z, z')] +2  C(u',z) D_z  \delta(z, z')\right)\,.
\la{PCbra}
\ee 

As anticipated in Section \ref{sec:asy-cond}, in the limit $u\to -\infty$ the bracket \eqref{PCbra} diverges. As explained there, this divergence can be eliminated by defining the corner charge aspects \eqref{charge-aspects}, whose subleading component is
\be\la{hatp}
p(z) :=\lim_{u\to-\infty} \frac{8}{\kappa^2}\f12\left(\cP(u,z)- uD\cM_\C(u,z)\right) \,.
\ee
Then using \eqref{PCbra} and \eqref{CM},
\begin{empheq}[box=\fbox]{align} \la{PC}
 \{p(z), C(u', z')\} =& - \f{u'}2 D_z\left[\left(\pa_{u'}C(u',z) - 2 D_z^2\right)\delta(z, z')\right]\cr
&-\left(   C(u',z) D_z  \delta(z, z')+\f12 D_{z}[C(u',z)\delta(z, z')]\right)\,.
 \end{empheq}
For later convenience, we can use the general formula
\be\la{form2}
f(z) D_z^s \delta(z,z')=\sum_{n=0}^s(-1)^n \f{s!}{n!(s-n)!}(D^nf)(z') D_z^{s-n}\delta(z,z')\,,
\ee
to
rewrite the RHS of \eqref{PC} in terms of $C(u',z')$ as 
\be\la{PC2}
 \{p(z), C(u', z')\} = u' D^3_z \delta(z, z')
-\f12 (u'\pa_{u'} +3)C(u',z') D_z\delta(z, z') + D_{z'}C(u',{{z'}}) \delta(z, z')\,. 
\ee
The asymptotic holomorphic super-Lorentz  charges and their symmetry action take the form \cite{Kapec:2014opa}
\be
Q_{Y}:= \int_S  Y(z)p(z),\qquad \delta^1_Y C(u,z):= \{ Q_Y, C(u, z)\}\,.
\ee
Together with \eqref{PC2}, one finds
\be\la{QYC}
 \delta^1_Y C(u,z)&= 
 \f{u}2 \left(\delta^0_{DY} C\right)(u, z) +   Y(z) D_{z}C(u,z) + \f32C(u,z) D_{z}Y(z)\,.
\ee
The $Y$ transformation  can be written in terms of the Lie derivative action
\be 
\label{Lie-action-subleading}
Y(z) D C(u,z) + \f32C(u,z) D Y(z)
= m^Am^B \left[\left( \cL_Y  - \frac12 DY\right) C_{AB}\right]\,,
\ee
which precisely agrees with the super-Lorentz transformations on the shear \cite{Kapec:2014opa, Compere:2018ylh, Freidel:2021yqe}. Note that the factor of 1/2 in \eqref{hatp} ensures that the charge action \eqref{QYC} reproduces the Lie derivative action \eqref{Lie-action-subleading} with the correct normalization. This explains the prefactor of the subleading corner aspect introduced in \eqref{corq}, and implies that the charge  $Q_Y$ is associated with a vector field
\be
\xi_Y=  \frac{u}2 DY\pa_u + Y m\,.
\ee

\subsection{Sub-subleading action}\la{sec:c-subsub}

Similarly, the sub-subleading charge aspect takes the form
\be
\label{sub-sub-T}
\cT(u,z)=   \int_{+\infty}^u \rd u'  \dot{\cT}(u',z)\,.
\ee
As before, equation \eqref{T} suggests the decomposition
\be\la{DT}
\cT= \cT_{\mathrm{SSS}} +\cT_{\mathrm{SSH}}+\cT_{\mathrm{SHH}}+\cT_{\mathrm{HHS}}+\cT_{\mathrm{HHH}}\,,
\ee
where the first three terms on the RHS correspond respectively to the first three terms in \eqref{T}, while the last two terms arise from the last term in \eqref{T} upon splitting $\mathcal{M}_{\mathbb{C}}$ according to \eqref{DM}.
We obtain the following  brackets (see Appendix \ref{app:comm} for details)
\be
\{ \cT_{\mathrm{HHH}}(u,z), C(u', z')\}
&=- \frac3{16} \kappa^2  \pa_{u'}\left[ C(u',z)  \left(\int_u^{u'} \rd u'' C(u'',z)   \right) \theta(u'-u)\right]  \delta(z, z')\,,
\la{Thhh}\\
\{ \cT_{\mathrm{HHS}}(u,z), C(u', z')\}
&= \f38 \kappa^2 \left(\int_u^{u'} \rd u'' C(u'',z)   \right)\theta(u'-u) D_z^2\delta(z, z')\,,
\\
\{\cT_{\mathrm{SHH}}(u,z), C(u', z')\}
&= \f{\kappa^2}4  (u'-u)\theta(u'-u)  D_z\left[ C(u',z) D_z \delta(z, z') \right] \,,
\\
\{ \cT_{\mathrm{SSH}}(u,z), C(u', z')\}
&= \f{\kappa^2}8\pa_{u'}\left(  D_z^2[ C(u',z) \delta(z, z') ]  \frac{(u'-u)^2}{2} \theta(u'-u) \right)\,,
\\
\{ \cT_{\mathrm{SSS}}(u,z), C(u', z')\}
&=-\f{\kappa^2}8 (u'-u)^2  \theta(u'-u)  D_z^4\delta(z, z') \,.
\ee

Putting everything together, these imply the following bracket of \eqref{sub-sub-T} with $C$ for $u'>u$  
\be 
\f8{\kappa^2} \{  \cT(u,z), C(u', z')\} &= \frac{(u'-u)^2}{2} D_z^2\left[\left(\pa_{u'}C(u',z) - 2 D_z^2\right)\delta(z, z')\right] \cr
&+2(u'-u) D_z\left[C(u',z)D_z\delta(z, z')+\f12 D_z[ C(u',z) \delta(z, z') ]\right ]\cr
&  -  \f32\left(\int_u^{u'}\rd u'' C(u'', z)\right) \left[\left(\pa_{u'}C(u',z) - 2 D_z^2\right)\delta(z, z')\right] \cr
&- \frac32 [C(u',z)]^2 \delta(z, z')\,  .
\ee
We see again that the generator  admitting a well defined limit $u\to -\infty$ is not 
$\cT(u,z)$ but the renormalized charge aspect defined in \eqref{corq}, namely\footnote{One uses that $
\frac{(u'-u)^2}{2}+u(u'-u) +\frac{u^2}{2}= \frac{{u'}^2}{2}$. One also uses our boundary condition $C\to0$ when $u\to +\infty$.}
\be
\label{sssq}
t(z):= \lim_{u\to -\infty} \frac{8}{\kappa^2}\f13\left(\cT(u,z) - u D\cP(u,z)+ \frac{u^2}{2}D^2 \cM_\C(u,z)
{ -\f32 \left(\int_{+\infty}^u \!\! C\right)}\cM_\C \right)\,. 
\ee
This combination is then such that 
\begin{empheq}[box=\fbox]{align}
\label{tC}
 \{ t(z), C(u', z')\} &= \left(\frac{{u'}^2}{6} D_z^2 -\frac12 \left(\int_{+\infty}^{u'} \!\! C(z)\right) \right) \left[\left(\pa_{u'}C(u',z) - 2 D_z^2\right)\delta(z, z')\right] \cr
&+\f23 u' D_z \left[C(u',z)D_z\delta(z, z')+\f12 D_z[ C(u',z) \delta(z, z') ]\right ]\cr
&- \f12  [C(u',z)]^2 \delta(z, z').
\end{empheq}
This symmetry action distinguishes itself from the leading and subleading ones through the following features:
1) it is non-local in time and 2) it involves collinear excitations.
It will be convenient to break down this symmetry action into a soft, hard and collinear component, $t=t_{\rm{S}}+ t_{\rm{H}}+t_{\rm{C}}$, where the collinear terms are cubic in $C,\bar{C}$.
As before, upon introducing the spin-2 charge and the spin-2 variation
 \be\label{QZ}
Q_Z :=\f8{\kappa^2}\int  Z(z) t( z), \qquad \delta^2_Z C(u, z)  := \{Q_Z,  C(u, z)\}\,,
 \ee 
 \eqref{tC} allows us to conclude that
 \be \label{TransQ}
 \delta^2_Z C(u,z) &= - \f{u^2}6 \left(\delta^0_{ D^2 Z } C \right) + \f23 u \left(\delta^1_{ D Z } C \right)  + D^2\left[ Z \left(\int_{+\infty}^u\!\!C\right)\right]  -\frac{Z}2\left( C^2 + \p_u C \left(\int_{+\infty}^u\!\!C\right) \right). 
 \ee 
The last term is a non-linear transformation consisting of products of $C(u,z)$ at possibly different times, but evaluated at the same point on the celestial sphere. We will investigate this collinear component in Section \ref{sec:corrections}.
 
 The linear transformation associated with the first two terms in \eqref{TransQ} can be described by introducing the notion of pseudo-vector fields.
 Vector fields on $\scri^{+}$ are linear combinations over the space of functions on $\scri^{+}$ of the individual vector fields $m$, $\bar{m}$ and $\p_u$ which act as differential operators on $\mathcal{I}^+$. 
 Holomorphic vectors only involve linear combinations of $(m,\pa_u)$. 
 Note that according to our discussion in Section \ref{sec:EOM}, the vectors  $(m,\pa_u)$ were assigned dimension $1$. We also introduce the pseudo-differential operator $\pa_u^{-1}$ of dimension $-1$.\footnote{This is the space-time version of the dimension-lowering operators encountered in sub-subleading conformally soft theorems \cite{Guevara:2019ypd, Pate:2019lpp}.} A holomorphic pseudo-vector of spin $s$ is then given by the product  $D_s=m^s\pa_u^{1-s}$. These pseudo-vectors are of dimension $1$ and they satisfy a generalization of the Leibniz rule \cite{DickeyB}
 \be 
\pa_u^{1-s} F = \sum_{n=0}^\infty \frac{(1-s)_n}{n!} (\pa_u^n F)\, \pa_u^{1 -s -n},
\ee
where $(x)_n = x(x - 1)...(x - n + 1)$ is the falling factorial.

 A pseudo-vector field is simply a linear combination over $\scri^+$
 of pseudo-vectors.  The set of pseudo-vectors is naturally equipped with a Poisson bracket which is defined as the commutator of pseudo-vector fields restricted to dimension  one\footnote{ It can be shown that the commutator of two holomorphic pseudo-vectors also contains a sum of terms proportional to $\pa_u^{-n} D_{s}$, which are of lower dimension. }
 \be 
 \{ F_s D_s , G_{s'} D_{s'} \}= F_s (D_s G_{s'}) D_{s'} - G_{s'} (D_{s'} F_s) D_{s}.
 \ee

We now see that the linear component of the transformation  \eqref{TransQ} is associated with the pseudo-vector field
 \be\la{xiZ}
\xi_Z = Z D_2  +\f23 u D Z D_1  +
\f{u^2}6 D^2 Z D_0,
\ee
where $D_0=\pa_u$ and $D_1= m$ are standard vectors while $D_2 = m^2 \pa_u^{-1}$ is a pseudo-vector. Again, the  factor of 1/3 in \eqref{sssq} is to ensure the correct normalization of the $u$-independent contribution in \eqref{xiZ}.

For later convenience, we can again use \eqref{form2} to rewrite the RHS of \eqref{tC} in terms of $C(u',z')$ as
\be \la{tC2} 
\begin{split}
 \{ t(z), C(u', z')\} &= -\frac{u'^2}{3}D_z^4\delta(z,z') \\
&+   \frac{1}{6}\left(\left[u'^2\p_{u'}^2 + 6 u'\p_{u'} + 6 \right]  \int_{+\infty}^{u'}du'' C(u'',z')\right) D_z^2\delta(z, z')\\
&- \frac{2}{3}\left(\left[u'\p_{u'} + 3 \right] D_{z'} \int_{+\infty}^{u'}du'' C(u'',z')\right) D_z\delta(z,z') \\
&+ \delta(z,z') D_{z'}^2\int_{+\infty}^{u'}du'' C(u'',z') \cr & - \f12 \pa_{u'}\left( C(u',z') \int_{+\infty}^{u'}du'' C(u'',z')\right) \delta(z, z'). 
\end{split}
\ee
As we will demonstrate in the next section, upon Fourier transforming, this symmetry action (excluding the non-linear contribution in the last line) can be recast into the sub-subleading soft theorem \cite{Cachazo:2014fwa, Campiglia:2016efb, Pate:2019lpp}.


\section{From conservation laws to soft theorems}
\label{cl-st}
In this section we demonstrate that the leading, subleading and sub-subleading soft graviton theorems follow from conservation laws associated with the charges \eqref{corq}. The analysis at the leading and subleading orders was first done in \cite{Strominger:2013jfa, Kapec:2014opa} and is presented herein for completeness. The derivation of the sub-subleading soft graviton theorem as a consequence of conservation of the charges \eqref{sssq} is new. 

Amplitudes in gravity\footnote{This behaviour is universal at tree-level.  In the quantum theory, the leading soft theorem remains universal, but $S^{(1)}$ and $S^{(2)}$ may receive one- and two-loop exact corrections.} have universal behavior in the limit when one of the gravitons becomes soft \cite{Weinberg:1965nx, Cachazo:2014fwa}. In particular,
\be 
 \langle {\rm out}|a_{\pm }(\omega \hat{q}) \mathcal{S} |{\rm in}\rangle = \left(S^{(0)}_{\pm} + S^{(1)}_{\pm} + S^{(2)}_{\pm} \right) \langle {\rm out}| \mathcal{S}|{\rm in} \rangle + \mathcal{O}(\omega^2).
\ee
Here $S^{(i)}_{\pm}$ for $i = 0, 1 , 2$ are the leading, subleading and sub-subleading soft factors \cite{Weinberg:1965nx, Cachazo:2014fwa},
\be 
S^{(0)}_{\pm} &= \frac{\kappa}{2}\sum_{k = 1}^n \frac{(p_k\cdot \varepsilon^{\pm})^2}{p_k \cdot q},\la{llsf}\\
S^{(1)}_{\pm} &= -\frac{i\kappa}{2}\sum_{k = 1}^n \frac{(p_k\cdot \varepsilon^{\pm})(q\cdot J_k \cdot\varepsilon^{\pm})}{p_k \cdot q},\la{ssf}\\
S^{(2)}_{\pm} &= -\frac{\kappa}{4} \sum_{k = 1}^n \frac{(\varepsilon^{\pm} \cdot J_k \cdot q)^2}{p_k \cdot q},\la{sssf}
\ee
where the subscripts refer to the helicity of the soft graviton. The (outgoing) graviton has momentum $q = \omega\hat{q}$ and polarization
\be  
\varepsilon^{\pm \pm}_{\mu\nu} = \varepsilon^{\pm}_{\mu} \varepsilon^{\pm}_{\nu},
\ee
while $p_k$ and $J_k$ are the momenta and angular momenta of all other (hard) particles.

Following \cite{Strominger:2013jfa}, the large-$r$ mode expansion\footnote{Note that our definition of $C= C_{AB} m^A m^B$  already includes the polarization factors.} of $C$ near $\mathcal{I}^+$ is
\be 
\la{mode-exp}
C(u, \hat{x}) = \frac{i\kappa}{8\pi^2}  \int_0^{\infty} d\omega \left[a_-^{\rm out\dagger}(\omega \hat{x}) e^{i\omega u} - a_{+}^{\rm out}(\omega\hat{x}) e^{-i\omega u} \right].
\ee
$\bar{C}$ takes a similar form related to \eqref{mode-exp} by Hermitian conjugation. 
One can check (see Appendix \ref{app:conventions}) that the commutators \eqref{NCC} imply the standard commutation relations for the modes,
\be \label{comm}
 [a_{\pm}(\omega \hat{x}), a_{\pm}^\dagger(\omega' \hat{x}')] &=(2\pi)^3  \frac{2}{\omega}  \delta(\omega-\omega')\delta(z,z').
\ee

We also define the Fourier modes \cite{Strominger:2013jfa}
\be 
\label{Nmode}
\begin{split}
N^{\omega} := \int_{-\infty}^{\infty} du e^{i\omega u} \p_u \bar C. 
\end{split}
\ee
This notation is consistent with the previous sections where $N$ was defined as the variable \textit{conjugate} to $C$ which is $\p_u \bar{C}$. As such, $N^{\omega}$ are the Fourier modes of \textit{negative} helicity gravitons.

Leading, subleading and sub-subleading negative-helicity (outgoing) soft gravitons correspond to \cite{Kapec:2016jld, Campiglia:2016efb}
\begin{subequations}\label{lsss}
\be 
N^{(0)} &= \frac{1}{2}\lim_{\omega \rightarrow 0^+} \left(N^{\omega} + N^{-\omega} \right) = -\frac{\kappa}{8\pi}  \lim_{\omega \rightarrow 0^+}\omega\left(a_{ +}^{\rm out \dagger}(\omega \hat{x}) + a_{ -}^{\rm out}(\omega\hat{x}) \right),\la{N0} \\
N^{(1)} &= -\frac{i}{2}\lim_{\omega \rightarrow 0^+}\p_{\omega}\left(N^{\omega} - N^{-\omega} \right) = { -}\frac{i\kappa}{8\pi}  \lim_{\omega \rightarrow 0^+}(1+\omega\p_{\omega})\left(a_{ +}^{\rm out \dagger}(\omega \hat{x}) - a_{ -}^{\rm out}(\omega  \hat{x}) \right),\la{N1}\\
N^{(2)} &=  -\frac{1}{4}\lim_{\omega \rightarrow 0^+}\p_{\omega}^2(N^{\omega} + N^{-\omega}) = \frac{\kappa}{16\pi}  \lim_{\omega \rightarrow 0^+} \p_{\omega}(1+\omega\p_{\omega})  \left(a_{ +}^{\rm out \dagger}(\omega \hat{x}) + a_{ -}^{\rm out}(\omega  \hat{x}) \right)\,.\la{N2}
\ee
\end{subequations}
Using \eqref{Nmode}, \eqref{lsss} can be equivalently written as
\be 
\label{soft-modes}
N^{(0)}= \int_{-\infty}^{\infty} du N, \quad N^{(1)} = \int_{-\infty}^{\infty} du u N , \quad N^{(2)} = \frac{1}{2}\int_{-\infty}^{\infty} du u^2 N,
\ee
which can in turn be related to the soft charges\footnote{Here the label S refers to the contributions to $\cM_\C,  \cP,  \cT$ linear in $N$ (i.e. containing no hard modes) in the decompositions \eqref{DM}, \eqref{DP}, \eqref{DT}.}
\begin{subequations}
\be
m_{\rm S}(z)&= \frac{8}{\kappa^2}\f12  \lim_{u\to -\infty}  D^2  \p_u^{-1} N(u, z)\,,\la{msoft}\\
p_{\rm S}(z)&= \frac{8}{\kappa^2}\f14 \lim_{u\to -\infty} D^3\left[\p_u^{-2} N(u, z)- u \p_u^{-1} N(u, z) \right]\,,\la{psoft}\\
t_{\rm S}(z)&=  \frac{8}{\kappa^2} \f16 \lim_{u\to -\infty} D^4\left[\p_u^{-3} N(u, z) - u \p_u^{-2} N(u, z)+ \frac{u^2}{2}\p_u^{-1} N(u, z) \right]\,. \la{tsoft}
\ee
\end{subequations}
The Leibniz rule \eqref{Leib} implies that
\be
\pa_{u}^{-1}\left( \frac{u^k}{k!} F(u)\right)&= 
(-1)^k\sum_{n=0}^{k}\frac{(-u)^{n}}{n!}
\pa_{u}^{-(1+k-n)  }F(u)\,,\la{Leibn}
\ee
which allows us to rewrite the subleading and sub-subleading renormalized soft charge aspects in terms of the  soft modes as 
\be\la{qsoft}
m_{\rm S}(z)=- \frac{4}{\kappa^2}    D^2 N^{(0)}(z)\,, \qquad 
p_{\rm S}(z)=  \frac{2}{\kappa^2}   D^3 N^{(1)}(z)\,,\qquad
t_{\rm S}(z)=- \frac{4}{3\kappa^2}   D^4 N^{(2)}(z)\,.
\ee
The analogous relations for $\bN=N_{AB}m^A m^B$  
are obtained by Hermitian conjugation.

The transformation properties of $C$ under the symmetries \eqref{corq} derived in the previous sections can be converted into actions of hard charges on asymptotic Fock states by means of inverting the mode expansion \eqref{mode-exp},
\be
\label{C-fourier}
\widetilde{C}(\omega,z):=
\int_{-\infty}^{\infty}\rd u e^{i\omega u} C(u, z) &= 
\frac{i\kappa}{4\pi}
\left(a_-^{\rm out\dagger}(-\omega \hat{x})\theta(-\omega) - a_{+}^{\rm out}(\omega\hat{x})\theta(\omega) \right) \,.
\ee
In particular, one finds \eqref{mC}, \eqref{PC} and \eqref{tC} imply the following commutators\footnote{Here the label H refers to the contributions to $\hat{q}$ in \eqref{corq} that are neither soft nor collinear.} 
\be
[m_{\mathbb{C} {\rm H}}(z), a_+^{{\rm out}}(\omega\hat{x}') ] &= -  \omega a_+^{\rm out}(\omega \hat{x}') \delta(z,z'),\la{l-hard-action} \\ 
[p_{\rm H}(z), a_+^{\rm out}(\omega \hat{x}')] &=  i \left(h_+ D_z\delta(z, z') -\delta(z, z') D_{z'} \right) a_+^{\rm out}(\omega\hat{x}') , \la{sl-hard-action}\\
[t_{\rm H}(z), a^{{\rm out} }_+(\omega \hat{x}')] &=  \frac{1}{6}\Big(2h_+(2h_+ - 1)D_z^2\delta(z,z') \cr 
&\hspace{30pt} - 8h_+ D_z\delta(z,z') D_{z'}  + 6\delta(z,z') D_{z'}^2 \Big) \omega^{-1} a^{{\rm out}}_+(\omega\hat{x}') + \mathcal{O}(\kappa^3),\la{sll-hard-action}\ee
where $\omega >0$
and we defined the left-moving conformal weights\footnote{These become diagonal in a conformal primary basis \cite{Kapec:2016jld, Pasterski:2016qvg}.}
\be \la{weights}
2h_{\pm} = -\omega\p_{\omega} \pm 2
\ee
associated with positive $(+)$ and negative $(-)$ helicity gravitons respectively. 
For the last commutator, we used (see Appendix \ref{app:conventions})
\be 
\label{FT-piC}
\widetilde{\pa_u^{-1} C}(\omega,z)&= \frac{\widetilde{C}(\omega,z)}{-i\omega -\epsilon} .
\ee
The commutators with negative helicity modes are implied by brackets of the charges with $\bar{C}$ and can be shown to take the same form with $h_+ \rightarrow h_-$. Similar mode expansions apply near $\mathcal{I}^-$ \cite{He:2014laa} allowing for commutators of the charges with the incoming modes to be computed. 
The $\mathcal{O}(\kappa^3)$ contribution arises from the quadratic term in \eqref{tC}. We postpone the analysis of this correction to Section \ref{sec:corrections}.

Without loss of generality, in the following sections we set $P = 1$ in which case the celestial sphere is flattened to a plane. All formulas can be covariantized by simply replacing $\p_z$ by $D_z$. We spell out our conventions in Appendix \ref{app:conventions}.

\subsection{Leading soft theorem}

  Using the parameterizations \eqref{mompar} of the momenta, the leading soft factor \eqref{llsf} becomes 
\be 
\label{par-lsf}
\begin{split}
S^{(0)}_+ &= -\frac{\kappa}{2\omega}\sum_{k =1}^n { \epsilon_k\omega_k}\frac{\bz - \bz_k}{z - z_k}, \quad
S^{(0)}_{-} = -\frac{\kappa}{2\omega}\sum_{k =1}^n { \epsilon_k\omega_k} \frac{z - z_k}{\bz - \bz_k}.
\end{split}
\ee
$\epsilon_k = \pm 1$ distinguishes between incoming $(-)$ and outgoing $(+)$ particles. For a negative helicity insertion we have
\be 
\label{wi}
\p_z^2 S^{(0)}_{-} = -\frac{\pi \kappa}{{ \omega}}\sum_{k =1}^n{ \epsilon_k \omega_k} \delta^{(2)}(z - z_k).
\ee

The matching condition\footnote{More precisely, $\left. m_{\mathbb{C}}(z) \right|_{\mathcal{I}^+_-} = \left. m_{\mathbb{C}}(\epsilon(z)) \right|_{\mathcal{I}^-_+}$
where $\epsilon(z)$ is the inversion $z \rightarrow -\frac{1}{\bz}, \bz \rightarrow -\frac{1}{z}$. To avoid clutter, we follow \cite{Strominger:2013jfa} and take the inversion to be implicit in all matching relations. } 
\be 
\left. m_{\mathbb{C}}(z) \right|_{\mathcal{I}^+_-} = \left. m_{\mathbb{C}}(z) \right|_{\mathcal{I}^-_+},
\ee
where the transverse coordinates $z$ at $\mathcal{I}^+_-$ and $\mathcal{I}^-_+$ are antipodally related \cite{Strominger:2013jfa}, then implies the conservation law
\be 
\la{lcl}
\langle{\rm out}|\left. m_{\mathbb{C}}(z) \right|_{\mathcal{I}^+_-} \mathcal{S} - \mathcal{S}  \left. m_{\mathbb{C}}(z) \right|_{\mathcal{I}^-_+} |{\rm in}\rangle = 0.
\ee
$m_{\mathbb{C}}$ consists of a soft component, namely a leading soft graviton \eqref{qsoft} and a hard component whose action on asymptotic states is implied by the commutator \eqref{l-hard-action}. 
We find
\be 
\frac{1}{\pi \kappa}\lim_{\omega \rightarrow 0} \omega \p_z^2 \langle {\rm out}|a^{{\rm out}}_-(\omega \hat{x}) \mathcal{S}|{\rm in} \rangle + \sum_{k = 1}^n{ \epsilon_k\omega_k}\delta^{(2)}(z - z_k) \langle {\rm out}| \mathcal{S}|{\rm in} \rangle = 0\,,
\ee
where the first term is associated with the soft part and the second term arises from the action of the hard part on asymptotic states.\footnote{One assumes that  $m_{\mathbb{C} {\rm H}}(z) |0 \rangle=0$ which can be achieved by normal ordering.} We have used crossing symmetry 
\be 
\la{crossing}
\langle{\rm out}|a_-^{\rm out}(\omega\hat{x})\mathcal{S}|{\rm in}\rangle =  \langle{\rm out}|\mathcal{S} a_+^{\rm in \dagger}(\omega \hat{x})|{\rm in}\rangle
\ee
to rewrite the soft charge in terms of an outgoing soft insertion. 
Rearranging, we recover \eqref{wi}. Note that at $O(\omega^{s - 1})$, \eqref{crossing} implies
\be 
\lim_{\omega \rightarrow 0^+} \p_{\omega}^s \left(\omega\langle{\rm out}| a_-^{{\rm out}}(\omega \hat{x}) S|{\rm in}\rangle\right) = (-1)^{s+1}\lim_{\omega \rightarrow 0^+} \p_{\omega}^s \left(\omega\langle {\rm out}|S a^{{\rm in}\dagger}_+(-\omega\hat{x})|{\rm out}\rangle\right). 
\ee
Therefore, the soft component of \eqref{lcl} and its subleading counterparts below will be twice that of an outgoing soft insertion \cite{He:2014laa, Kapec:2014opa}.

\subsection{Subleading soft theorem}
In the parameterization \eqref{mompar}, the subleading soft factor \eqref{ssf} becomes \cite{Kapec:2014opa}
\be 
\label{par-ssf}
\begin{split}
S^{(1)}_+ &= \frac{\kappa}{2}\sum_{k = 1}^n \frac{(\bz - \bz_k)^2}{z - z_k}\left[\frac{2\bar{h}_k}{\bz - \bz_k} - \p_{\bz_k} \right], \quad
S^{(1)}_- = \frac{\kappa}{2}\sum_{k = 1}^n \frac{(z - z_k)^2}{\bz - \bz_k}\left[\frac{2h_k}{z - z_k} - \p_{z_k} \right],
\end{split}
\ee
or equivalently, for a negative helicity soft insertion,
\be 
\label{sst}
\p_z^3 S_-^{(1)} = 2\pi \kappa \sum_{k = 1}^n \left[h_k \p_{z_k}\delta^{(2)}(z - z_k) -  \delta^{(2)}(z - z_k)\p_{z_k} \right]\,.
\ee
Here 
\be \la{weights1}
2h_{k} = -\omega_k\p_{\omega_k} + s_k, \quad 2\bar h_{k} = -\omega_k\p_{\omega_k} - s_k,
\ee
where $s_k$ are graviton helicities.

As before, imposing an antipodal matching condition on $p(z)$
\be 
\left.p(z) \right|_{\mathcal{I}^+_-} = \left.p(z) \right|_{\mathcal{I}^-_+},
\ee
upon splitting $p(z)$ according to \eqref{P} and using \eqref{qsoft} to identify the soft component with a subleading soft insertion and \eqref{sl-hard-action} to compute the action of the hard component on asymptotic states, we find \cite{Kapec:2014opa}
\be 
 \frac{i}{2 \kappa \pi} \p_z^3 \lim_{\omega \rightarrow 0}(1 + \omega\p_{\omega}) \langle{\rm out}|a^{{{\rm {out}}}}_-(\omega\hat{x})\mathcal{S} |{\rm in}\rangle - \sum_{k = 1}^n i \left(h_k \p_z \delta^{(2)}(z - z_k) - \delta^{(2)}(z- z_k) \p_{z_k} \right) \langle{\rm out}|\mathcal{S}|{\rm in}\rangle = 0.
\ee
As expected, this agrees with \eqref{sst}.

\subsection{Sub-subleading soft theorem}
\label{sec:subsub}

Using the standard parameterization for the momenta spelled out in Appendix \ref{app:conventions}, the sub-subleading soft factor \eqref{sssf} can be put into the form \cite{Campiglia:2016jdj, Campiglia:2016efb, Pate:2019lpp} 
\be
\label{mellsfc}
\begin{split}
S^{(2)}_- = -\frac{\kappa \omega}{4} \sum_{k = 1}^n \frac{z - z_k}{\bz - \bz_k}\left[2h_k(2h_k - 1) -  2(z - z_k) 2h_k \p_{z_k} + (z - z_k)^2\p^2_{z_k}\right](\epsilon_k \omega_k)^{-1} ,\\ 
S^{(2)}_+ = -\frac{\kappa \omega}{4} \sum_{k = 1}^n \frac{\bz - \bz_k}{z - z_k}\left[2\bar{h}_k(2\bar{h}_k - 1)-  2(\bz - \bz_k) 2\bar{h}_k \p_{\bz_k} + (\bz - \bz_k)^2\p^2_{\bz_k}\right](\epsilon_k \omega_k)^{-1}.
\end{split}
\ee

As before, focusing on the negative helicity insertion and taking $\p_z^4 S_-^{(2)}$, all terms localize to delta functions or derivatives thereof, namely
\be 
\label{final}
\begin{split}
\p_z^4 S_-^{(2)} = -\frac{\pi \kappa \omega}{2}  \sum_{k = 1}^n\Big(2h_k(2h_k - 1) \p_z^2\delta^{(2)}(z - z_k) - 4(2h_k) &\p_z\delta^{(2)}(z - z_k)\p_{z_k} \\
&+ 6 \delta^{(2)}(z - z_k)\p_{z_k}^2 \Big)(\epsilon_k\omega_k)^{-1}.
\end{split}
\ee
An antipodal matching condition for $t(z)$, 
\be 
\left.t(z)\right|_{\mathcal{I}^+_-} =\left. t(z)\right|_{\mathcal{I}^-_+}
\ee
then implies the conservation law
\be 
\label{cl}
\langle {\rm out}|\left. t(z)\right|_{\mathcal{I}^+_-}\mathcal{S} - \mathcal{S}\left. t(z)\right|_{\mathcal{I}^-_+}|{\rm in} \rangle = 0.
\ee
\eqref{qsoft} identifies the soft component of \eqref{cl} with a sub-subleading soft insertion
\be 
\begin{split}
\langle {\rm out}|\left. t_{\rm S}(z)\right|_{\mathcal{I}^+_-} \cS - \cS\left. t_{\rm S}(z)\right|_{\mathcal{I}^-_+}|{\rm in}\rangle = -\frac{1}{6 \kappa \pi}\lim_{\omega \rightarrow 0}\p_{\omega}(1 + \omega\p_{\omega})\p_{z}^4\left(\langle {\rm out}|a_{-}^{\rm out }(\omega\hat{x})\mathcal{S}|{\rm in}\rangle \right).
\end{split}
\ee
On the other hand, \eqref{sll-hard-action} allows for the hard component of \eqref{cl} to be written as
\be 
\begin{split}
\langle {\rm out}|\left. t_{\rm H}(z)\right|_{\mathcal{I}^+_-} \cS - \cS\left. t_{\rm H}(z)\right|_{\mathcal{I}^-_+}|{\rm in}\rangle =&- \frac{1}{6} \sum_{k = 1}^n\Big(2h_k(2h_k - 1)\p_z^2\delta^{(2)}(z -z_k)
- 8h_k \p_z\delta^{(2)}(z -z_k) \p_{z_k} \\
&+ 6\delta^{(2)}(z -z_k) \p_{z_k}^2 \Big)(\epsilon_k\omega_k)^{-1} \langle {\rm out}|S|{\rm in}\rangle + \mathcal{O}(\kappa^3).
\end{split}
\ee
The $\mathcal{O}(\kappa^3)$ contributions arise from the terms cubic in the fields whose action on asymptotic states will be computed using \eqref{Thhh} in Section \ref{sec:corrections}. 

Putting everything together, we find that to leading order in $\kappa$
\be 
\begin{split}
\lim_{\omega \rightarrow 0}&\p_{\omega}(1 + \omega\p_{\omega}) \p_{z}^4\left(\langle {\rm out}|a_{-}^{\rm out }(\omega\hat{x})\mathcal{S}|{\rm in}\rangle \right)   = 2 \p_z^4 S^{(2)} \langle {\rm out}|\mathcal{S}|{\rm in}\rangle =\\
&= -\pi \kappa \sum_{k = 1}^n \Big(2h_k(2h_k - 1) \p_z^2\delta^{(2)}(z - z_k) - 8h_k \p_z\delta^{(2)}(z - z_k)\p_{z_k} \cr
&\hspace{200pt}+ 6 \delta^{(2)}(z - z_k)\p_{z_k}^2 \Big)(\epsilon_k\omega_k)^{-1} \langle {\rm out}|\mathcal{S}|{\rm in}\rangle, \\
\end{split}
\ee
which remarkably agrees with \eqref{final}.

\section{Collinear contribution to the soft theorem}\la{sec:corrections}

In this section we consider the quadratic contribution to symmetry action \eqref{tC2}, which upon promoting the bracket to a commutator takes the form 
\be \la{t3C} 
 [ t_{\rm C}(z), C(u', z')]  =
 \f{i}2 \pa_{u}\left( C(u,z')\pa_u^{-1}C(u,z') \right) \delta(z, z').
\ee  
Upon Fourier transforming and using \eqref{C-fourier}, we find
\be 
\label{collinear-comm}
 [ t_{\rm C}(z), a_+^{\rm out}(\omega\hat{x}')]  & = -\f{2\pi}{\kappa}
 \widetilde{\left[\pa_u[ C \pa_u^{-1} C]\right]}(\omega) \delta(z, z')\cr 
 & = -\frac{ \omega}{\kappa} 
\int_{-\infty}^{+\infty}\rd \omega'\left(   \frac{ \widetilde{C}(\omega-\omega')\widetilde{C}(\omega')}{(\omega-\omega') -i\epsilon} \right) \delta(z, z'),
\ee
where we have used the convolution theorem and \eqref{FT-piC}.
Using \eqref{C-fourier} the RHS of \eqref{collinear-comm} can be re-expressed in terms of modes, namely
\be
 [ t_{\rm C}(z), a_+^{{\rm out}}(\omega\hat{x}')  ]
 &= \frac{\kappa \omega  }{ 16\pi^2} 
 \int_{0}^{\omega}\rd \omega'
 \left[  
\frac{ a_+((\omega - \omega')\hat{x}') a_+(\omega'\hat{x}') }{\omega - \omega'- i\epsilon}\right]
\delta(z, z')\cr
&-\frac{\kappa \omega^2  }{ 16\pi^2}  
 \int_{\omega}^{+\infty}\rd \omega'
  \f{\left[  
 a_+^\dagger ((\omega' - \omega)\hat{x}') a_+(\omega'\hat{x}') \right]}{(\omega - \omega'-i\epsilon)( \omega'-i\epsilon)}
\delta(z, z')\,.
\ee
Details are given in Appendix \ref{osc}. The commutators with the opposite helicity as well as incoming modes can be found similarly. 

Both terms represent collinear corrections to the sub-subleading soft theorem that can be traced back to classical, non-perturbative gravitational effects. The first term is a particle creating contribution which can be evaluated using the universal behavior of equal helicity gravitons in the collinear limit \cite{Bern:1998sv} which for positive helicity gravitons takes the form
\be 
\lim_{\hat{p}_i \rightarrow \hat{p}_j} \langle \omega_i \hat{p}_i; \omega_j \hat{p}_j|\mathcal{S}|{\rm in}\rangle = -\frac{\kappa}{2}\frac{\bz_{ij}}{z_{ij}} \frac{(\omega_i + \omega_j)^2}{\omega_i \omega_j}\langle  (\omega_i + \omega_j) \hat{p}_j|\mathcal{S}|{\rm in}\rangle + \cdots,
\ee
where $\cdots$ denote subleading terms in the collinear limit. 
The associated correction then becomes
\be 
\langle {\rm{out}} |\left. t_{\rm C}(z)\right|_{\mathcal{I}^+_-}\mathcal{S} &- \mathcal{S} \left. t_{\rm C}(z)\right|_{\mathcal{I}^-_+}|{\rm in}\rangle \cr
&\propto \kappa^2 \sum_{i} \omega_i \int_0^{\omega_i} \rd\omega' \frac{\omega_i^2}{\omega'(\omega' - \omega_i)^2}\delta(z,z_i) \langle \omega_i\hat{p}_i|\mathcal{S}|{\rm in} \rangle + \cdots, \\
&\propto \kappa^2 \lim_{\epsilon \rightarrow 0} \sum_{i} \omega_i B(\epsilon, -1 + \epsilon) \delta(z,z_i) \langle \omega_i\hat{p}_i |\mathcal{S}|{\rm in} \rangle + \cdots,
\ee
{where $B(\epsilon,-1+ \epsilon) = \frac{1}{\epsilon}$ is an Euler beta function resulting from the regulated integral.}
The second term can be shown to give rise to corrections with $\delta$-function support of the form
\be 
\langle {\rm{out}} |\left. t_{\rm C}(z)\right|_{\mathcal{I}^+_-}\mathcal{S} &- \mathcal{S} \left. t_{\rm C}(z)\right|_{\mathcal{I}^-_+}|{\rm in}\rangle \cr
&\propto \kappa \sum_{i,j} F_{ij}(\omega_i, \omega_j) \delta(z_i, z_j)\delta(z,z_i) \langle (\omega_i + \omega_j)\hat{p}_i|\mathcal{S}|{\rm in}\rangle + \cdots, 
\ee
where $F_{ij}$ are (possibly vanishing) functions of the external energies. We expect a careful treatment of contact terms in the original proof of \cite{Cachazo:2014fwa} to reveal these corrections.\footnote{We thank Freddy Cachazo for a discussion on this point.} We leave a complete understanding of their amplitudes origin and implications to future work.

\section{Conclusions}\la{sec:conc}

In this work we have established a clear connection between the spin-$2$ conservation equation and the sub-subleading soft theorem. We have learned that the non-linear nature of Einstein's equations  manifests itself in the sub-subleading soft theorem through collinear corrections. We have also revealed that, unlike the spin-$0$ and -$1$ symmetries responsible for the leading and subleading soft theorems,  the spin-$2$ symmetry is not simply an asymptotic diffeomorphism.
It involves a non-local transformation represented by a pseudo-vector field acting on $\scri$. The extension of this symmetry to the bulk of spacetime  remains mysterious to us. 
This result now puts us in a position to understand the nature of the spin-$2$ memory effect, which is a question we expect to return to in the near future.

From the S-matrix point of view, it is expected that this spin-$2$ charge, or rather its quadratic truncation, is one of the  canonical generators for the  $w_{1+ \infty}$ symmetry unraveled by celestial holography \cite{Guevara:2021abz, Strominger:2021lvk}. Interestingly, the connection between W$_N$ algebras, pseudo-vectors and  integrable systems has already been explored in the past \cite{Bakas:1989um}.
This suggests that there could be an exciting connection between asymptotic Einstein's equations and symmetries of integrable systems.

Finally, it is natural to wonder whether the collection of celestial Ward identities associated with the entire $w_{1+ \infty}$ symmetry tower possesses a gravitational dynamical interpretation.   We started to address this question in \cite{Freidel:2021ytz}, where 
first evidence for a canonical realization of a $w_{1+ \infty}$ loop algebra on the gravitational phase space was given.


\section*{Acknowledgements}

We would like to thank Freddy Cachazo and Sabrina Pasterski for helpful discussions and insights. We also thank  Andrew Strominger for feedback on a final version of this manuscript. 
Research at Perimeter Institute is supported in part by the Government of Canada through the Department of Innovation, Science and Economic Development Canada and by the Province of Ontario through the Ministry of Colleges and Universities. This project has received funding from the European Union's Horizon 2020 research and innovation programme under the Marie Sklodowska-Curie grant agreement No. 841923. The research of A.R. is additionally supported by the Stephen Hawking fellowship. 

\appendix
\section{Conformal derivative}\label{confD}

In this appendix we study the properties of the conformal derivative introduced in Section \ref{sec:EOM}. 
From the transformation \eqref{delq} of the metric we deduce that
\be
\label{gamma-var}
\delta_W \Gamma_{ AB }^C= - D_{ A}W \delta_{B}^C- D_{ B}W \delta_{A}^C +q_{AB}D^C W\,.
\ee
Let us assume that $O_{\langle A_1 \cdots A_s\rangle}$ is a spin $s$ field of dimension $\Delta$. We have 
\bea 
\delta_W D_{\langle A_0}O_{ A_1 \cdots A_s\rangle}
&=&
D_{\langle A_0}\delta_W O_{ A_1 \cdots A_s\rangle}
-\sum_{i=1}^s \delta_W \Gamma_{\langle A_0A_i}^{B_i}O_{ A_1 \cdots |B_i| \cdots A_s\rangle}\cr&=& 
D_{\langle A_0}( (\Delta-s) W O_{ A_1 \cdots A_s\rangle})
+2s  D_{\langle A_0}W O_{A_1 \cdots A_s\rangle} \cr 
&=& (\Delta -s) W 
D_{\langle A_0} O_{ A_1 \cdots A_s\rangle}) 
+ (\Delta + s)D_{\langle A_0}W O_{A_1 \cdots A_s\rangle}.
\eea  
Hence, by using the transformation \eqref{dUp} of the conformal connection $\Upsilon_A$, we see that
\be 
\cD_{\langle A_0}O_{ A_1 \cdots A_s\rangle}:= 
D_{\langle A_0}O_{ A_1 \cdots A_s\rangle} + 
(\Delta + s) \Upsilon_{\langle A_0}O_{ A_1 \cdots A_s\rangle}
\ee 
transforms covariantly.  
Similarly for a spin $-s$ field, one finds
\bea 
\delta_W D_{ A_0}O^{ \langle A_1 \cdots A_s\rangle}
&=& (\Delta +s) W 
D_{ A_0}O^{ \langle A_1 \cdots A_s\rangle} 
+ (\Delta - s)(D_{ A_0}W) O^{ \langle A_1 \cdots A_s\rangle}.
\eea 

We can also compute the transformation of
 \bea 
 \delta_W D_{\langle A} \Upsilon_{B\rangle}&=&
  D_{\langle A} \delta_W \Upsilon_{B\rangle} - 
 \delta_W \Gamma_{\langle A B\rangle}^C \Upsilon_{C}\cr 
 &=& - D_{\langle A} D_{B\rangle} W + 
 2 D_{\langle A} W \Upsilon_{B\rangle},\cr 
 \delta_W \Upsilon_{\langle A}\Upsilon_{B\rangle} &=& -2 D_{\langle A} W \Upsilon_{B\rangle}.
 \eea 
This implies that 
\be
\delta_W \tau_{AB} = 2 \delta_W (D_{\langle A} \Upsilon_{B\rangle} + \Upsilon_{\langle A}\Upsilon_{B\rangle}  ) =  -2 D_{\langle A} D_{B\rangle} W.
\ee 
This coincides with the transformation of $N_{AB}$ and hence
\be
\delta_W (N_{AB}-\tau_{AB}) =0.
\ee 
Since $C^{AB}$ is a primary field of $(\Delta, s) = (1,-2)$, we deduce that 
$\cD_B C^{AB}$ and $\cD_A\cD_B C^{AB}$ are respectively of dimension $(2,-1)$ and $(3,0)$. Moreover, 
\be 
\cD_A \cD_B C^{AB} &=
(D_A+ \Upsilon_A) (D_B -\Upsilon_B) C^{AB}\cr 
&= D_A D_B C^{AB} - (D_A \Upsilon_B +\Upsilon_A\Upsilon_B) C^{AB}\cr 
&= D_A D_B C^{AB} - \frac12  C_{AB}\tau^{AB}\,.
\ee 
Similarly, 
\be 
\cD_A \cD_B \tilde C^{AB} &=
D_A D_B \tilde{C}^{AB} - \frac12 \tilde{C}_{AB} \tau^{AB}\,,
\ee 
where $\tilde{C}^{AB}:= \epsilon^A{}_C C^{CB}$.
We can thus write the dual covariant mass as \cite{Freidel:2021qpz}
\be 
\tilde{\cM} &= \frac14 D_A D_B \tilde{C}^{AB} - \frac18 \tilde{C}_{AB} N^{AB}\cr 
&= \f14\cD_A \cD_B \tilde{C}^{AB} - \frac18 \tilde{C}_{AB} \hat{N}^{AB}.
\ee

In order to extend the definition of the conformally covariant derivative beyond the case of symmetric, traceless tensors, we use \eqref{gamma-var}
to note that for a spin-1 field $O_A$ of dimension $\Delta$ we have
\be
\delta_W D_{B}O_A&= 
D_{B}\d_W O_A-\d_W \Gamma^C_{AB} O_C\cr
&=(\Delta-1)D_{B} (WO_A)
+ D_{ A}W O_B+ D_{ B}W O_A -q_{AB} D^C W O_C\cr
&=(\Delta-1) WD_{B} O_A+ (\Delta-1)D_{ B}W O_A
+ 2 D_{ \langle A}W O_{B\rangle}
\,.
\ee
Hence, we see that
\be
\cD_B O_A&:= D_{B}O_A +(\Delta-1) \Upsilon_{ B}O_{ A}
+2\Upsilon_{ \langle A}O_{ B\rangle}
\ee
transforms covariantly as
\be
\d_W \cD_B O_A = (\Delta-1) W\cD_B O_A\,.
\ee
For a general field of spin $s$ and dimension $\Delta$, the conformal covariant derivative is given by
\be\la{WC}
 \cD_B O_{A_1\cdots A_s}:= D_B O_{A_1\cdots A_s}+(\Delta-s) \Upsilon_{ B}O_{A_1\cdots A_s}
 +2\sum_{i=1}^s \Upsilon_{ \langle B}O_{ |A_1\cdots |A_i\rangle\cdots A_s }\,.
\ee
Finally, we see that  the sphere metric $q_{AB}$, which is a field of spin $s=2$ and dimension $\Delta=0$, 
is compatible with the conformal connection, namely
\be
\cD_C q_{AB}&= -2 \Upsilon_C q_{AB}+2 \Upsilon_{ \langle C} q_{A\rangle B}+2 \Upsilon_{ \langle C} q_{B\rangle A}
=0\,.
\ee

\subsection{Commutators and curvature}
\label{comm-curv}

By means of the general formula \eqref{WC}, we can evaluate the commutator of the  Weyl-covariant derivative on a field of spin $1$ and dimension $\Delta$
\be 
[\cD_A , \cD_B ] V_C&=\cD_A (D_BV_C +(\Delta-1) \Upsilon_{ B} V_C +2\Upsilon_{ \langle B}V_{ C\rangle}) -A\leftrightarrow B\cr
&=D_A D_BV_C+(\Delta-1) (D_A \Upsilon_{ B}) V_C +2(D_A \Upsilon_{ \langle B})V_{ C\rangle}
-A\leftrightarrow B\cr 
&=[D_A, D_B]V_C +2\Delta (D_{[A} \Upsilon_{ B]}) V_C
+2(D_{[A{|}} \Upsilon_{  C})V_{{|} B]}
-2q_{C[B} (D_{A]} \Upsilon_{  D})V^{ D}\,.
\ee
Writing the conformal connection as $\Upsilon_A = D_A \varphi$ and working in complex coordinates, we can derive the identity
\be
(D_{[A{|}} \Upsilon_{  C})V_{{|} B]}
-q_{C[B} (D_{A]} \Upsilon_{  D})V^{ D}= q_{A[C}q_{D]B}V^D D_E \Upsilon^E\,.
\ee
Moreover, we can write the commutator of the  Weyl-covariant derivative as
\be
[\cD_A , \cD_B ] V_C&= {\cal R} q_{A[C}q_{D]B} V^D\,,
\ee
where, in analogy to \cite{Campiglia:2020qvc}, we have defined 
\be
{\cal R}:= R+2D_A \Upsilon^A\,,
\ee
and we recall $R$ denotes the scalar curvature of the 2-sphere metric $q_{AB}$.

Finally,
\be 
\delta_W D_A \tau^{AB}& =
D_A \delta_W \tau^{AB} +\delta_W \Gamma_{AC}^A \tau^{CB} +\delta_W \Gamma_{AC}^B\tau^{AC} \cr 
& =
4 D_A(W \tau^{AB}) - 2D_A D^{\langle A} D^{B\rangle }W  - 4 D_C W \tau^{CB} \cr 
&=4 W (D_A \tau^{AB}+\f14 D^BR) -  D^B( R  W +  \hat{\Delta} W),
\ee 
where $\hat\Delta=D_C D^C$ and $[\hat\Delta,  D^B] W= \dfrac{R}{2} D^B W$. We also have that 
\be 
\delta_W R&= 2 W R +2 \hat{\Delta} W\,,
\ee
from which
\be 
\delta_W (D^B R) = 2W D^B R + D^B(2 W R +2 \hat{\Delta} W)\,.
\ee
Therefore,
\be 
\delta_W(\frac{1}{2}D^B R + D_{A} \tau^{AB}) = 4 W (D_A \tau^{AB} + \frac{1}{2} D^B R).
\ee
This implies that $2 D_A \tau^{AB}+ D^B R$ is a primary field of dimension/spin $(3,-1)$. If we assume it vanishes for one choice of the conformal orbit, it vanishes at all times.

\subsection{Spin connection and variations}\label{SpinC}
 The spin connection $\Omega$ appears in 
\be 
D_A m_B -D_B m_A &=  \Omega \epsilon_{AB}.
\ee 
Using \eqref{trm} and contracting with $m^{A} \bar{m}^B$, as well as with $\bar{m}^{A} m^B$, one finds
\be 
i\Omega = \bar{m}^A D m_A = \bar{m}^A \bar{D} m_A. 
\ee
Moreover,
\be
D_A m^A &= (m^A\bar{m}^B + \bar{m}^A m^B) D_A m_B = \bar{m}^B D m_B = i\Omega.
\ee 
We can expand the vector field as $Y^A = Ym^A + \bar{Y}\bar{m}^A$ where $Y=Y^A \bar{m}_A$. 
This means that, under GBMS transformations ($W = \frac{1}{2} D_A Y^A$), we have
\be
\delta_{(T,Y)} m_A
&= Y^B D_B m_A + m_B D_A Y^B- \frac12 D_B Y^B m_A \cr 
&=i\Omega (Y m_A -\bar{Y}\bar{m}_A)+  D_A \bar{Y}- \frac12((D+i\Omega)Y+(\bar{D}-i\Omega)\bar{Y} ) m_A \,, \\
m^A \delta_{(T,Y)} m_A
&= (D- i\Omega)\bar{Y}\,  ,\\
 \bar{m}^A \delta_{(T,Y)} m_A
 &=
   \frac12 ( (\bar{D}+i\Omega) \bar{Y} -(D-i\Omega) Y )\,,\\
  \frac12 q^{AB}\delta_{(T,Y)} q_{AB} &= m^A \bar{m}^B \delta_{(T,Y)}(\bar{m}_A m_B) + \bar{m}^A m^B \delta_{(T,Y)}(\bar{m}_A m_B)\cr &= 
  \bar{m}^A \delta_{(T,Y)} m_A + m^A \delta_{(T,Y)} \bar{m}_A=0.
\ee

Then for a primary spin $s$ operator, 
\be 
O_{A_1\cdots A_s} &= O_s \bar{m}_{A_1}\cdots  \bar{m}_{A_s} + \bar{O}_s m_{A_1}\cdots  m_{A_s},\cr 
O^{A_1\cdots A_s} &= O_{-s}m^{A_1}\cdots  m^{A_s}  + \bar{O}_{-s}  \bar{m}^{A_1}\cdots  \bar{m}^{A_s},
\ee
where $O_{-s}= \bar{O}_{s}$, one finds the variations
\be 
m^{A_1}\cdots  m^{A_s}\delta_{Y}O_{A_1\cdots A_s}
&= D_Y O_s + s O_s(m^A D_Y \bar{m}_A) + s D Y^B \bar{m}_B  O_{s} + (\Delta -s) W_Y O_{s} \cr 
&=(YD+ \bar{Y} \bar{D}) O_s -i  s \Omega (Y+\bar{Y}){O_s} + s O_{s} (D Y + Y \bar{m}_BDm^B  ) \cr 
&+ \frac12 O_{s} (\Delta -s)  ((D+i\Omega)Y+(\bar{D}-i\Omega)\bar{Y}) \cr 
&=Y[(D-is\Omega)O_s]   + \bar{Y} [(\bar{D}-is\Omega)O_s]  \cr 
&+ \frac12 (\Delta +s)  O_{s} [(D+i\Omega)Y]+ \frac12 (\Delta -s)O_s[(\bar{D}-i\Omega)\bar{Y}]\,.
\la{dYO}
\\
O_{A_1\cdots A_s} \delta_{Y} m^{A_1}\cdots  m^{A_s}&=   -\frac{s}2 O_s[(D-i\Omega) Y]+ \frac{s}2 O_s [(\bar{D}+i\Omega) \bar{Y} ]\,.
\ee 
We thus see that the general transformation \eqref{dYO} is consistent with the shear transformation \eqref{Lie-action-subleading}
(recall that $C_{AB}$ is a primary of $(\Delta,s)=(1,2)$ and footnote \ref{f1}).
Finally, we can write
\be
\delta_{Y} O_s&=
(Y D+\bar{Y} \bar{D}) O_s  
+ \frac \Delta 2 O_{s} (D+i\Omega)Y  
+ \frac \Delta2 O_s (\bar{D}-i\Omega)\bar{Y}\,.
\ee

\section{Charge commutators}
\label{app:comm}

In this appendix we spell out the steps leading to the results presented in Section \ref{sec:commutators}.

\subsection{Leading charges}
The commutators of the soft and hard parts of $\mathcal{M}_{\mathbb{C}}$ with $C$ are
\bea 
\label{MsC}
\{\cM_{\mathrm{S}}(u,z), C(u', z')\}
&=& \int_{+\infty}^u \rd u''   D_z\{ \cJ(u'',z), C(u', z')\}\cr
&=& -\f{ \kappa^2}4\theta(u'-u) D_z^2\delta(z, z')\,,
\eea
and
\bea 
\label{MhC2}
\{ \cM_{\mathrm{H}}(u,z), C(u', z')\}
&=& \frac14 \int_{+\infty}^u \rd u''   C(u'',z)\{ \cN(u'',z), C(u', z')\}\cr
&=&  -\f{\kappa^2}8\int_{+\infty}^u \rd u''   C(u'',z) \pa_{u'} \delta(u'-u'') \delta(z, z')\cr
&=&\f{\kappa^2}8\pa_{u'}[ C(u',z) \theta(u'-u) ] \delta(z, z')\,,
\eea
where we have used
\be 
\int^u_{\infty} \rd u'' \delta(u'' - u') = -\int_u^{\infty}\rd u''\delta(u'' - u') = -\theta(u' - u),
\ee
with
\be 
\la{theta-fctn}
\theta(x) = \begin{cases} 0, x < 0, \\
1, x \geq 0.
\end{cases}
\ee
 
 \subsection{Subleading charges} 
\label{app:subleading-charges}
The commutators of the terms in $\mathcal{P}$ with $C$ are
\bea 
\{ \cP_{\mathrm{SS}}(u,z), C(u', z')\} &=& \int_{+\infty}^u \rd u''   D_z\{ \cM_{\mathrm{S}}(u'',z), C(u', z')\}\cr
&=& {{\frac{\kappa^2}{4}}} \int_u^{+\infty}\rd u''  \theta(u'-u'') D_z^3\delta(z, z')\cr
&=& {{\frac{\kappa^2}{4}}}(u'-u) \theta(u'-u)  D_z^3\delta(z, z')\,,
\eea
\bea
\{ \cP_{\mathrm{SH}}(u,z), C(u', z')\}
&=&  \int_{+\infty}^u \rd u''   D_z \{ \cM_{\mathrm{H}}(u'',z), C(u', z')\}\cr
&=& {{\frac{\kappa^2}{8}}}\p_{u'}\int_{+\infty}^u \rd u'' D_z\left(C(u',z')\theta(u' - u'')\delta(z,z')\right)\cr
&=&{{-\frac{\kappa^2}{8}}}D_z\pa_{u'}[ C(u',z) \theta(u'-u)\delta(z, z')(u' - u) ]\,,
\eea
and
\bea
\{ \cP_{\mathrm{HH}}(u,z), C(u', z')\}
&=&  \int_{+\infty}^u \rd u''   C(u'',z) \{ \cJ(u'',z), C(u', z')\}\cr
&=&  {{\frac{\kappa^2}{4}}} \int_{+\infty}^u \rd u''   C(u'',z)  \delta(u''-u') D_z \delta(z, z')\cr
&=&-{{\frac{\kappa^2}{4}}}[ C(u',z) \theta(u'-u) ] D_z\delta(z, z'),
\eea
where in the last bracket  we used \eqref{JC}. 

\subsection{Sub-subleading charges}
For the sub-subleading charge we find
\be 
\{ \cT_{\mathrm{HHH}}(u,z), C(u', z')\}
&= \frac32  \int_{+\infty}^u \rd u''   C(u'',z) \{ \cM_{H}(u'',z), C(u', z')\}\cr
&=   \frac3{16}\kappa^2   \int_{+\infty}^u \rd u''   C(u'',z)  \pa_{u'}[ C(u',z) \theta(u'-u'') ] \delta(z, z')\cr
&=- \frac3{16} \kappa^2  \pa_{u'}\left[ C(u',z)  \left(\int_u^{u'} \rd u'' C(u'',z)   \right) \theta(u'-u)\right]  \delta(z, z')\,,
\ee
\bea
\{ \cT_{\mathrm{HHS}}(u,z), C(u', z')\}
&=& \frac32  \int_{+\infty}^u \rd u''   C(u'',z) \{ \cM_{S}(u'',z), C(u', z')\}\cr
&=&  -\f38 \kappa^2  \int_{+\infty}^u \rd u''   C(u'',z)  \theta(u'-u'') D_z^2\delta(z, z')\cr
&=& \f38 \kappa^2 \left(\int_u^{u'} \rd u'' C(u'',z)   \right)\theta(u'-u) D_z^2\delta(z, z')\,,
\eea
\bea
\{\cT_{\mathrm{SHH}}(u,z), C(u', z')\}
&=&   \int_{+\infty}^u \rd u''  D_z \{\cP_{HH}(u'',z), C(u', z')\}\cr
&=& -\f{\kappa^2}4 \int_{+\infty}^u \rd u''  D_z\left[ C(u',z) D_z\delta(z, z') \right]  \theta(u'-u'') \cr
&=& \f{\kappa^2}4  (u'-u)\theta(u'-u)  D_z\left[ C(u',z) D_z \delta(z, z') \right] \,,
\eea
\bea
\{ \cT_{\mathrm{SSH}}(u,z), C(u', z')\}
&=&   \int_{+\infty}^u \rd u''  D_z \{ \cP_{SH}(u'',z), C(u', z')\}\cr
&=&  -\f{\kappa^2}8 \int_{+\infty}^u \rd u'' \pa_{u'}\left( D_z^2[ C(u',z) \delta(z, z') ]  (u'-u'')\theta(u'-u'') \right)\cr
&=& \f{\kappa^2}8\pa_{u'}\left(  D_z^2[ C(u',z) \delta(z, z') ]  \frac{(u'-u)^2}{2} \theta(u'-u) \right)\,,
\eea
and finally
\bea
\{ \cT_{\mathrm{SSS}}(u,z), C(u', z')\}
&=&   \int_{+\infty}^u \rd u''  D_z \{ \cP_{SS}(u'',z), C(u', z')\}\cr
&=&  \f{\kappa^2}4 \int_{+\infty}^u \rd u'' (u'-u'') \theta(u'-u'')  D_z^4\delta(z, z') \cr
&=&-\f{\kappa^2}8 (u'-u)^2  \theta(u'-u)  D_z^4\delta(z, z') \,.
\eea

\section{Conventions}
\label{app:conventions}

To demonstrate the equivalence between symmetries and soft theorems in Section \ref{cl-st} it is convenient to work in flat retarded coordinates
\be
\begin{split}
x^{\mu} &= u \p_z\p_{\bz} \hat{q}^{\mu}(z, \bz) + r \hat{q}^{\mu}(z,\bz), \\
\hat{q}^{\mu}(z, \bz) &= \frac{1}{\sqrt{2}}\left(1 + z\bz, z + \bz, -i(z - \bz), 1- z\bz \right) {:= \hat{x}(z, \bz)},
\end{split}
\ee
in which the Minkowski metric becomes
	\be \rd s^2 = \rd x^{\mu}\rd x_{\mu} = -2\rd u\rd r+ 2r^2\rd z\rd\bz,\ee
and the celestial sphere is conformally mapped to a plane. This corresponds to choosing $P = 1$ in \eqref{trm}.

More generally, we parameterize Cartesian coordinate massless 4-momenta as
	\be \label{mompar}
	\begin{split}
		p_k^{\mu} 
		&= \frac{\epsilon_k \omega_k}{\sqrt{2}}(1 + z_k \bz_k, z_k + \bz_k, -i(z_k - \bz_k), 1 - z_k\bz_k),\\
		q &= \frac{\omega}{\sqrt{2}}(1 + z \bz, z + \bz, -i(z - \bz), 1 - z\bz),
		\end{split}
	\ee
	with $\mu=0,1,2,3$ and $\epsilon_k = \pm 1$ for outgoing and incoming momenta respectively. $z_k$ is the spatial location at which a particle of momentum $p_k$ crosses ${\cal I}^+$.  In this parameterization,
	\be\label{pot} p_1\cdot p_2=-\epsilon_1 \epsilon_2 \omega_1 \omega_2 z_{12}\bz_{12},\ee
	where \be z_{12}=z_1-z_2, ~~~ \bz_{12}=\bz_1-\bz_2.\ee 
It can be shown \cite{He:2014laa, Kapec:2014opa, Kapec:2016jld, Campiglia:2016efb, Pate:2019lpp} that in these coordinates, the soft factors \eqref{llsf}-\eqref{sssf} indeed take the forms \eqref{par-lsf}, \eqref{par-ssf} and \eqref{mellsfc}.

The mode expansions of the shear and the news take the form \cite{He:2014laa} (see \cite{Strominger:2017zoo} for a review) 
\be 
\begin{split}
C(u, \hat{x}) &= \frac{i\kappa}{8\pi^2} \int_0^{\infty} \rd\omega \left[a_-^{\rm out\dagger}(\omega \hat{x}) e^{i\omega u} - a_{+}^{\rm out}(\omega\hat{x}) e^{-i\omega u} \right],\\
\hat N(u, \hat{x}) &= -\frac{\kappa}{8\pi^2} \int_0^{\infty} \rd\omega \omega \left[a_+^{\rm out\dagger}(\omega \hat{x}) e^{i\omega u} + a_{-}^{\rm out}(\omega\hat{x}) e^{-i\omega u} \right],
\end{split}
\ee
where $\hat{N}:=\pa_u \bar{C}.$
From this we have for $\omega > 0$
\be 
\omega a_-(\omega \hat{x}) &= -\frac{4\pi}{\kappa} \int \rd u \hat N(u,\hat{x}) e^{i\omega u},\cr 
a_-^\dagger(\omega' \hat{x}') &= -\frac{4i\pi}{\kappa} \int \rd u' C(u',\hat{x}') e^{-i\omega' u'}.
\ee 
We now recall \eqref{NCC},
\be 
[ \hat{N}(u,z), C(u', z') ] &= {-i} \f\k2 \delta(u-u') \delta(z,z')\,,
\ee
which allows us to compute
\be 
\omega [a_-(\omega \hat{x}), a_-^\dagger(\omega' \hat{x}')] &= i \left( \f{4\pi}{\k}\right)^2 
\int \rd u \rd u' [N(u, \hat{x}),C(u',\hat{x}')]  e^{i\omega u} e^{-i\omega' u'} \cr 
&= 16 \pi^3 \delta(\omega-\omega')\delta(z,z').
\ee 

Moreover, considering the Fourier transform of $C$,
\be 
\label{C-modes}
\widetilde{C}(\omega, z)&= \int_{-\infty}^{\infty} \rd u e^{i\omega u} C(u, \hat{x}) =  \frac{i\kappa}{4\pi} \left[a_-^{\rm out \dagger}(-\omega\hat{x}) \theta(-\omega) -  a_+^{\rm out}(\omega\hat{x}) \theta(\omega) \right],
\ee
we find that
\be 
(\pa_u^{-1} C) (u,z) &:= \int_{+\infty}^u \rd u' C(u',z) = \f1{2\pi}\int_{-\infty}^{+\infty} \rd \omega  \widetilde{C}(\omega,z) \int_{+\infty}^u \rd u' e^{-i(\omega-i\epsilon) u'}\cr 
& = \f{1}{2i\pi} \int_{-\infty}^{+\infty}\rd \omega \f{\widetilde{C}(\omega,z)}{-\omega + i\epsilon } e^{-i(\omega-i\epsilon) u},
\ee
or equivalently
\be 
\la{FTpui}
\widetilde{\pa_u^{-1} C}(\omega,z)=\frac{ \widetilde{C}(\omega,z)}{-i\omega - \epsilon }. 
\ee

\section{Collinear terms}\label{osc}

\eqref{FTpui} and the convolution theorem imply that
\be 
\widetilde{\left[\pa_u[ C \pa_u^{-1} C]\right]}(\omega, z)= \frac{\omega}{2\pi} \int_{-\infty}^{\infty} \rd\omega' \widetilde{C}(\omega', z) \frac{\widetilde{C}(\omega - \omega', z)}{(\omega - \omega') - i\epsilon}.
\ee

We can use \eqref{t3C} to evaluate the commutator of the cubic contribution to the sub-subleading charge with an annihilation operator. We find
\be 
\frac{\omega}{2\pi} \int \rd\omega' \widetilde{C}(\omega', z') \frac{\widetilde{C}(\omega - \omega', z')}{(\omega - \omega') - i\epsilon} &= -\frac{\omega \kappa^2}{(4\pi)^2 2 \pi}\int \rd\omega' \Big[\frac{a_+^{\rm out}(\omega'\hat{x}')a_+^{\rm out}((\omega - \omega')\hat{x}')}{\omega - \omega' - i\epsilon}\theta(\omega')\theta(\omega - \omega')
\cr
&-\frac{\omega a_-^{{\rm out}\dagger}((\omega' - \omega)\hat{x}') a_+^{\rm out}(\omega'\hat{x}')}{(\omega' - i\epsilon)(\omega - \omega' - i\epsilon)} \theta(\omega') \theta(\omega' - \omega) \Big]
\ee
and therefore
\be 
 [ t_{\rm C}(z), a^{\rm out}_+(\omega\hat{x}')  ] &= -\frac{\omega}{\kappa} \int \rd\omega' \widetilde{C}(\omega',z') \frac{\widetilde{C}(\omega - \omega',z')}{(\omega - \omega') - i\epsilon}\delta(z, z') \cr
 &= \frac{\kappa\omega}{(4\pi)^2}\int_0^{\omega} \rd\omega'  \frac{a_+^{\rm out}(\omega'\hat{x}')a_+^{\rm out}(\omega - \omega')\hat{x}')}{\omega - \omega' - i\epsilon}\delta(z,z')\cr
&-\frac{\kappa\omega^2}{(4\pi)^2}\int_{\omega}^{\infty} \rd\omega' \frac{a_-^{{\rm out}\dagger}((\omega' - \omega)\hat{x}') a_+^{\rm out}(\omega'\hat{x}')}{(\omega' - i\epsilon)(\omega - \omega' - i\epsilon)}\delta(z,z'), \quad \omega >0.
\ee

\bibliographystyle{bib-style2.bst}
\bibliography{biblio-fluxes}

\end{document}